\documentclass[11pt, a4paper]{article}
\pdfoutput=1

\usepackage[usenames, dvipsnames]{color}
\usepackage[T1]{fontenc}
\usepackage[latin1]{inputenc}
\usepackage{amsmath}
\usepackage[pdftex]{graphicx}
\usepackage{lastpage}
\usepackage{amssymb} 
\usepackage{hyperref}

\usepackage{cite}
\usepackage{soul}
\usepackage{mathtools}

\usepackage{subcaption}

\usepackage{esvect}
\renewcommand*\vec{\vv}

\usepackage{xspace}
\newcommand*{\ie}{i.e., }
\newcommand*{\eg}{e.g., }
\newcommand*{\fig}{Fig.\@\xspace}
\newcommand*{\Eq}{Eq.\@\xspace}

\usepackage{slashed}
\usepackage{braket}
\usepackage{simplewick}
\usepackage{bbm}
\newcommand\idop{\mathbbm 1}

\usepackage{subcaption}

\usepackage{titling}
\usepackage{authblk}

\newcommand{\ex}{\text{e}}
   
\newcommand\numberthis{\addtocounter{equation}{1}\tag{\theequation}}

\usepackage{xifthen}% Provides \isempty test
\newcommand*\diff{\mathrm{d}} % Straight differential
\newcommand*\ldiff[2][]{ \ifthenelse{\isempty{#1}}{ \diff #2}{\diff^#1#2} \,} % Differential with measure; the mandatory argument is the name of the measure, the option one is the dimension
\let\limitint\int % Only when I provide explicit limits for the integration, I need to do the spacing myself
\renewcommand{\int}{\limitint \!} % The standard integral should have correct spacing
\usepackage{breqn}

\title{Finding Critical States of Enhanced Memory Capacity\newline in Attractive Cold Bosons
	\\[1.5\baselineskip]}

\author[1,2,3]{Gia Dvali\thanks{georgi.dvali@physik.uni-muenchen.de}}
\author[1,2]{Marco Michel\thanks{marco.michel@physik.uni-muenchen.de}}
\author[1,2]{Sebastian Zell\thanks{sebastian.zell@campus.lmu.de}}
\affil[1]{Arnold Sommerfeld Center, Ludwig-Maximilians-Universit\"at, \mbox{Theresienstraße 37, 80333 M\"unchen, Germany}}
\affil[2]{Max-Planck-Institut für Physik, F\"ohringer Ring 6, 80805 M\"unchen, Germany}
\affil[3]{Center for Cosmology and Particle Physics, Department of Physics,\mbox{ New York University, 726 Broadway, New York, NY 10003, USA}}

\begin{document}

\allowdisplaybreaks

\maketitle
\begin{abstract}
 We discuss a class of quantum theories which exhibit a sharply increased memory storage capacity due to emergent gapless degrees of freedom. 
 Their realization, both theoretical and experimental, is of great interest. On the one hand, such systems are motivated from a quantum information point of view. On the other hand,  they can provide a framework for simulating systems with enhanced capacity of pattern storage, such as black holes and neural networks. In this paper, we develop an analytic method that enables us to find critical states with increased storage capabilities in a generic system of cold bosons with weak attractive interactions.  
 The enhancement of memory capacity arises  
 when the occupation number $N$ of certain modes
 reaches a critical level. Such modes,  via negative energy couplings, assist others in becoming effectively  gapless. This leads to 
 degenerate microstates labeled by the occupation numbers of the nearly-gapless modes.
 In the limit of large $N$,  they become 
 exactly gapless and their decoherence time diverges. In this way, a system becomes an ideal storer of quantum information.	
 We demonstrate our method on  a prototype model of $N$ attractive cold bosons contained in a one-dimensional box with Dirichlet boundary conditions. Although we limit ourselves to a truncated system, we observe a rich structure of quantum phases with a critical point of enhanced memory capacity.

\end{abstract}

\clearpage

\tableofcontents

\section{The Role and Importance of Gapless Modes
for Information Storage}
\subsection{Emergence of Gapless Modes in Attractive Bosonic Systems}

 A physical system is fully characterized by its degrees of freedom and by the rules of the interactions among them. 
 Each degree of freedom 
 corresponds to a particular oscillatory mode of the system.  
 In quantum field theory, such modes are described as quantum oscillators that can exist in various excited states. The level of excitation 
 of a mode $k$ in a given state  $\ket{n_k}$ can be conveniently described by an occupation  number $n_k$ of the corresponding quantum oscillator, with 
 the usual creation/annihilation operators
 $\hat{a}_k^{\dagger}$, $\hat{a}_k$ and 
 the number operator $\hat{n}_k = \hat{a}_k^{\dagger}\hat{a}_k$ (where $k =0,1,\ldots,K$). 
 We shall limit ourselves to bosonic degrees of freedom,  
 which satisfy the standard canonical commutation relations:    
 \begin{equation} 
 [\hat{a}_j,\hat{a}_k^{\dagger}] = \delta_{jk}\,,\qquad
 [\hat{a}_j,\hat{a}_k]  =   [\hat{a}_j^{\dagger},\hat{a}_k^{\dagger}] =0\,.   
 \label{algebra} 
 \end{equation} 
 One of the most important characteristics of a system is the energy level-spacing between states of different occupation numbers, 
 i.e., $\ket{n_k}$ and  $\ket{n_k \pm 1}$, which we shall denote by $E_k$.
 
We will view quantum states from a perspective of quantum information theory. 
When a degree of freedom can choose among $d$ different possible states $\ket{n_k}$ with $n_k = 0,1,\ldots,d-1$, it represents a qudit. Then information can be stored in the set of $d^{K+1}$
basic states $\ket{n_0,n_1,\ldots}$ and each basic state 
corresponds to a distinct pattern. In particular, we will be interested in the energy cost of information storage and read-out. In general, for a pattern $\ket{n_0,n_1,\ldots}$, it is given as $E_{n_0,n_1,\ldots} = \sum_k E_k n_k$. This means that the transition between patterns in which the occupation numbers differ significantly is in general expected to be costly.

   In this paper, we study systems in which information can be recorded and read out more efficiently. Adopting the criteria formulated in \cite{neural1, neural2},  we shall be interested in systems that exhibit the following properties: 
   	\begin{enumerate}
   		\item  A largest possible number of patterns can be stored  within a maximally narrow energy gap; and 
   		\item The stored patterns can be rearranged under the influence of the softest possible external stimuli.    
   	\end{enumerate} 

It will become evident that such a situation can be  achieved when the 
 modes that store patterns behave as either gapless or nearly-gapless. 
The latter term requires some quantification. Under a nearly-gapless 
mode we mean a mode 
 for which the minimal excitation energy $E_k$ is much less than the typical energy gap $\Delta E_{\text{typical}}$, expected for the system of a given size. For instance, for a non-relativistic particle of mass $M$ trapped  in a box of size 
$L$, one would expect the energy gap between the ground state and the first excited state to be set by the inverse size of the box, 
$\Delta E_{\text{typical}}  \sim {\hbar^2 \over 2 M L^2}$.

 As modes become nearly-gapless, $E_k \rightarrow0$, the energy cost of transitions between different states shrinks. Therefore, the density of patterns that can be stored in a given energy gap increases exponentially and such systems exhibit a sharply enhanced memory 
 storage capacity. 
 Moreover, in such a limit it becomes highly 
 energy-efficient to redial and/or  read out the
 stored patterns. 
 Indeed, the patterns can be rearranged under an influence of arbitrarily soft external stimuli \cite{neural1}. Hence, systems that feature gapless modes possess precisely those properties of efficient information processing that we are after.

 In this light, it is very important to understand  what physical mechanisms can allow a finite size system to deliver gapless modes.  
   We shall focus on a mechanism schematized in \cite{neural1,neural2},   
 which can be referred to as the phenomenon of {\it assisted}  gaplessness. 
 It represents a generalization of the original idea \cite{quantumPhase} 
of information storage in gapless modes that 
emerge in certain quantum critical states of attractive bosons.     
 The key principle of the assisted gaplessness mechanism  is easy to summarize.  
 If the interaction energy among the degrees of freedom of a system is negative, a high excitation of some of those modes lowers the excitation energy thresholds for the others.  In this way,  these highly excited degrees of freedom  play the role of {\it master}  modes that {\it assist} others 
 in becoming easily-excitable.  
 When the occupation numbers of 
 the master modes reach certain critical levels, the assisted modes become nearly-gapless. At this point, a state of  enhanced memory storage capacity is 
 attained.

  In order to understand the essence of the phenomenon, following \cite{neural1,neural2}, we consider an exemplary situation in which a mode $\hat{n}_0$, which is typically the one with the smallest kinetic energy $E_0$, can be highly occupied and has the following negative-energy coupling with a set of $K$ modes:
  \begin{equation} \label{FH1}
  \hat{H} =\sum_{k=1}^{K} E_k (1 -  \alpha \hat{n}_0) \hat{n}_k \,  +  E_0 \hat{n}_0 \, + \ldots \,.  
 \end{equation} 
 In this way, $\hat{n}_0$ becomes the {\it master} mode. 
 Its interaction energy with each mode 
$\hat{n}_k$ is proportional 
to the threshold energy $E_k$  of the latter modes 
via an universal proportionality  
constant $\alpha$. Due to the negative sign, such a 
connection is {\it excitatory}, \ie it is  energetically favorable to simultaneously excite the inter-coupled modes.  
 
\begin{figure}
 	\centering 
 	\begin{subfigure}{0.4\textwidth}
 		\includegraphics[width=\textwidth]{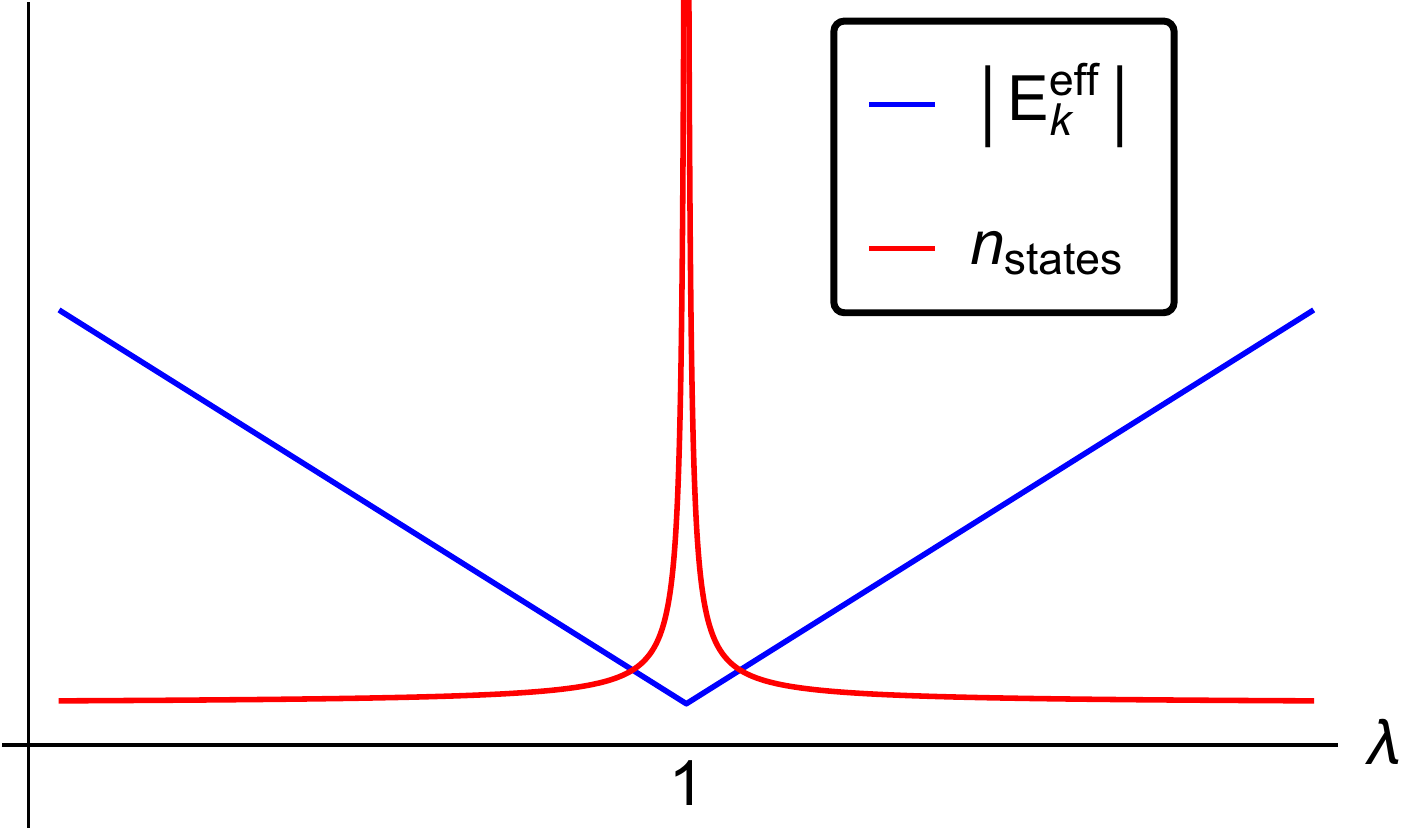}
 		\caption{Effective energy gap  $E_k^{\text{eff}}$ (as defined in \Eq \eqref{GAPH}) in blue and density of states $n_{\text{states}} \sim 1/E_k^{\text{eff}}$ in red. In the limit of infinite particle number, the energy gap shrinks to zero at the critical point $\lambda= 1$ and consequently the density of states diverges there.} 
 		\label{sfig:gapDensityCaroon}
 	\end{subfigure}
 	\hspace{0.1\textwidth}
 	\begin{subfigure}{0.4\textwidth}
 		\includegraphics[width=\textwidth]{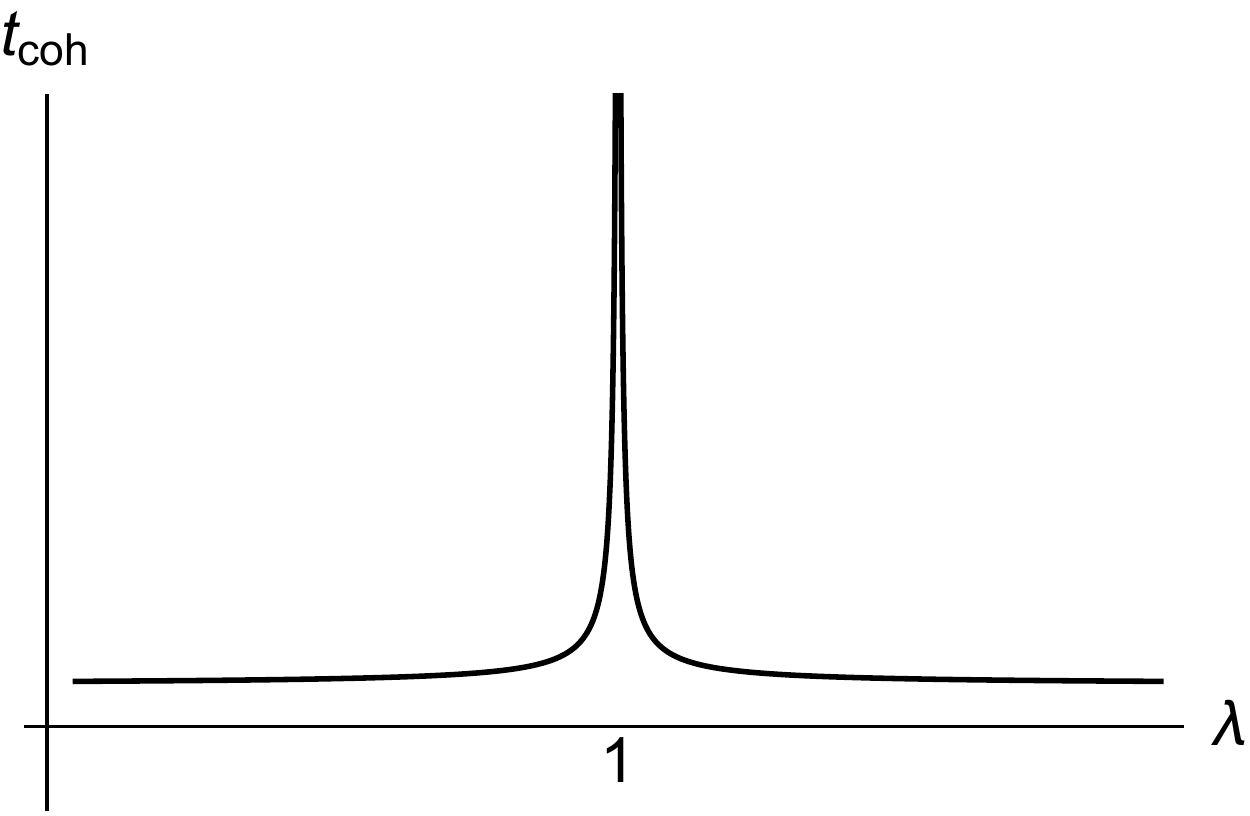}
 		\caption{Decoherence time of a typical state. In the limit of infinite particle number, it diverges at the critical point $\lambda= 1$.}
 		\label{sfig:decoherenceCartoon}
 	\end{subfigure}
 	\caption{Schematic plots of the behavior of a system  of attractive bosons such as \eqref{FH1} in the vicinity of a critical point with nearly-gapless modes that arise due to assisted gaplessness.}
 \end{figure}

 Thus, on states in which the occupation number 
 of the master mode is $\langle \hat{n}_0 \rangle = N_0 $,  the effective gap 
 for other modes is lowered as, 
  \begin{equation} \label{GAPH}
  E_k^{\text{eff}} = E_k (1 -  \lambda) \,,   
 \end{equation} 
 where we introduced the collective coupling
 \begin{equation}
 	\lambda := \alpha N_0 \,.
 	\end{equation}
Therefore, the master mode $\hat{n}_0$ assists 
the rest of the modes in becoming more easily excitable.  Accordingly, once the occupation number of the master  mode reaches a critical value $N_0 = \alpha^{-1}$ corresponding to $\lambda =1$, the  assisted modes $\hat{n}_k$ become {\it gapless}. The corresponding excitation energy as a function of $\lambda$ is plotted in \fig \ref{sfig:gapDensityCaroon}.
Note that small deviations $\delta N_0 \sim 1$ of the 
occupation number $N_0$ from the critical value result in the generation of an effective gap 
of order $ E_k^{\text{eff}}  \sim E_k \alpha$. Correspondingly, the smaller  $\alpha$ is, the less sensitive  the gap becomes to small fluctuations of the occupation number $N_0$ around its critical value.
Therefore, we will throughout focus on cases with small $\alpha$ and large $N_0$.   

In the model \eqref{FH1}, all states of the form $\ket{N_0,n_1,...n_K}$, where $n_{k\neq 0}$ can take different possible values from $0$ to $d-1$, become degenerate in energy. This means that the gapless modes can store an exponentially large number of patterns, ${\mathcal N_{pattern}} =d^K$, within an arbitrarily narrow energy gap. The resulting density of states as function of the collective coupling $\lambda$ is plotted in \fig \ref{sfig:gapDensityCaroon}.
  The neighborhood in the Fock space with a large number of states that fit within a narrow energy gap is characterized by the same macroscopic parameter, i.e., a macroscopically large occupation number $N_0$ of the $\hat{n}_0$-mode.  If $d$ is not large,  we can say that the states $\ket{N_0,n_1,...n_K}$  for all possible values of $n_{k\neq 0}$ are 
 macroscopically-indistinguishable. Hence,  they  form 
 a set of microstates belonging to the same macrostate.
 Correspondingly, we can define the microstate entropy as the  logarithm  of the number of such nearly-degenerate microstates, 
 \begin{equation} \label{ENTR}
 {\rm Entropy } = K \ln(d) \,. 
 \end{equation} 
 
 Apart from a low energy cost, systems that feature gapless modes possess a second property that makes them ideal information storers. Namely, gaplessness of a given mode 
 $\hat{b}$ implies that the disturbance from other modes is small
 since otherwise the gaplessness would be destroyed.  
 This suppression of interaction is either due to a relatively 
high energy gap
of the other modes or due to a very weak coupling to them (or both). 
In both cases, the time evolution into the other modes is small. 
Therefore, gaplessness implies a long information storage time.

This point can be illustrated in very general terms on an example with
one additional mode $\hat{c}$ that in particular may impersonate 
an environment. By assumption, both modes are approximate 
eigenmodes of the Hamiltonian. The latter can then be written in 
the following $2\times 2$ form:
  \begin{equation} \label{disturbanceHamiltonian}
  \hat{H}  = \begin{pmatrix}
    \hat{b}^{\dagger}   & \hat{c}^{\dagger}   \\  
\end{pmatrix}\begin{pmatrix}
  \epsilon    &  g  \\
    g  &  \epsilon'  
\end{pmatrix}\begin{pmatrix}
    \hat{b}    \\
    \hat{c}  
\end{pmatrix} \,.
\end{equation}
In order not to disturb the nearly-vanishing gap $\epsilon$, $g$ must satisfy  
 $g \ll \sqrt{\epsilon\epsilon'} $, \ie
$g$ must be sufficiently small in order to maintain the level-splitting.
This implies that either the time evolution will be strongly suppressed due to 
the large level-splitting (in the regime of $\epsilon'\gg \epsilon$) or 
the evolution timescale will be set by $t_{\text{coh}} \sim {\hbar \over \epsilon}$
and thus will be long (in the regime of  $\epsilon' \sim \epsilon$).
In both cases we have an effective protection of the stored information.
In other words, the information stored in a gapless mode 
$\hat{b}$ is maintained either due to the suppression of the amplitude of oscillations or due to a very long timescale of this transition.

 An important point of our analysis is that we do not introduce a gapless mode by hand but discover that such modes emerge even if the system 
 is confined within a box of finite size. This is a highly non-trivial 
 phenomenon that requires a critical balance between the 
 coupling and the occupation number.
 In this way, the decoherence time of quantum information stored in a gapless mode can be made arbitrarily long, even for fixed values of the size of the box and $\hbar$ (see \fig \ref{sfig:decoherenceCartoon}). 
 
As we have seen, the only requirements for the  the phenomenon of assisted gaplessness to take place in a bosonic system are a weak attractive interaction and a high occupation number of some of the modes. Therefore, as explained in \cite{neural1}, we expect that gapless modes, which lead to states of high memory storage capacity, are a generic phenomenon in these systems. Only the details, such as the exact number of the emergent gapless modes 
and the corresponding microstate entropy, depend on the symmetry structure and other details of the Hamiltonian.  

Interestingly, in the $D$-dimensional model of \cite{areaLaw},  the number of the emergent gapless modes and their microstate entropy 
scale as the area of a $D-1$-dimensional sphere, in similarity to the black hole Bekenstein entropy. These results suggest that systems with emergent gapless modes can offer a pathway for understanding the microscopic origin of black hole entropy and holography on simple prototype systems. In addition, such models serve as an existence proof of a non-gravitational system with area-law entropy and holography. We will elaborate on these points shortly in section \ref{ssec:simulation}.

  To summarize,  we expect that the phenomenon of emergence of gapless modes is a common phenomenon in bosonic systems with attractive interactions and its study is motivated 
from broad perspective of quantum information storage and processing, including practical applications,
\cite{quantumPhase, daniel, nico, goldstone,mischa1, mischa3}. Moreover, in the light of existing models \cite{areaLaw,neural2} 
that give an explicit microscopic description of the 
origin of the area-law of the microstate entropy and  
the emergence of holographic gapless modes, 
such systems can serve as a toy laboratory for
understanding analogous properties in black holes. 
  Finally, it has been suggested \cite{neural1,neural2} that a similar mechanism of the enhancement of  memory storage capacity can operate in neural networks, both quantum and classical.

In the present paper, we shall focus our study on the 
detailed understanding of the  phenomenon of assisted gaplessness.  We will provide a method that enables 
the search for states with gapless modes
in a systematic way in a theory of bosonic modes with a generic interaction Hamiltonian. The method relies on the Bogoliubov approximation, in which mode operators are replaced by their expectation values \cite{bogoliubov}. In this way, we transform the Hamiltonian in its Bogoliubov counterpart, which only depends on $c$-numbers.  We will show that it suffices to look for flat directions in this Bogoliubov Hamiltonian to conclude that gapless modes exist in the spectrum of the full quantum system. This allows us to  
replace the difficult problem of diagonalization 
of the Hamiltonian by a much simpler problem of extremizing
a $c$-number function. 

This approach, which we shall call $c$-number method, is a generalization of the one used in \cite{goldstone}, where it was shown for a specific model that the appearance of gapless modes at the critical point can be deduced from 
the minimization of a non-linear sigma-model obtained by replacing
the Hamiltonian by a $c$-number function.  
 Apart from its practical application, the $c$-number method also serves a second purpose. 
 The fact that any flat direction in the Bogoliubov Hamiltonian already implies a gapless mode supports the reasoning of \cite{neural1} that the emergence of gapless modes, which lead to states of high memory storage capacity, is rather generic, provided the interacting degrees of freedom of a system are bosonic and that some of the interaction energies are negative. 
 
After describing the $c$-number method in general, we apply it to a prototype model of attractive non-relativistic bosons in a one-dimensional box. To facilitate the analysis, we truncate the system to the three lowest momentum modes. We will show that this system of three interacting degrees of freedom possesses a critical point with a nearly-gapless mode.
The choice of our particular prototype model is mainly motivated by its presumed experimental simplicity.  The fact that the described phenomenon of memory enhancement  takes place already in such a simple system gives a natural hope that it could be experimentally studied under laboratory conditions, e.g., in a system of cold atoms. Such experiment would be highly interesting both from a general perspective of quantum information storage and due to the above mentioned connection to black holes and neural networks.
	
 The outline is as follows. In the remainder of this section, we elaborate on the potential of using bosons in a box to simulate other systems of enhanced memory storage such as black holes and neural networks. Moreover, we motivate the choice of our prototype system. In section \ref{sec:analysis}, we  present the $c$-number method for finding gapless modes in a generic system of attractive bosons. Section \ref{sec:3Mode} introduces our prototype model of 
  non-relativistic bosons in a one-dimensional box. In section \ref{sec:criticalLightModes}, we come to its critical point, at which a light mode appears and leads to the information  storage properties of our interest. We will first apply the $c$-number method to identify it and then confirm its existence by studying the full quantum system numerically. Moreover, we point out how  we could encode information in the occupation numbers of the light mode around the critical states by coupling the system to external degrees of freedom. In section \ref{sec:neural}, we discuss how we can map the properties of our model onto a neural network. We conclude in section \ref{sec:outlook} by summarizing our findings and emphasizing the importance of studying the critical point of a system of attractive cold bosons experimentally. Appendix \ref{app:formulas} contains some explicit formulas for our prototype systems and appendix \ref{app:periodic} is devoted to the exemplary application of the $c$-number method to a simpler model, which is analogous to our prototype model except for the choice of periodic boundary conditions and which has already been solved previously.
 
\subsection{Simulating Black Holes and Neural Networks}
 \label{ssec:simulation}

 Black holes are mysterious objects from the point of view of quantum information. One well-established  fact about them is that they 
 saturate the Bekenstein bound on information capacity
 \cite{bekensteinEntropy, bekensteinBound}.\footnote
 {On the order of magnitude, the Bekenstein bound coincides with the previously discovered limit on information capacity by Bremermann \cite{bremermannBoud}.}
  This means that for a given size of the system, the black hole is the state with maximal information, where its capacity of information storage is measured by the Bekenstein entropy  \cite{bekensteinEntropy}. This  entropy satisfies an area-law, i.e., it is given by the 
  area measured in units of the Planck scale.  
 Both facts are mysteries. 
 First, in order to accommodate such a high entropy, the black hole 
 must deliver qudits with extremely small energy gaps. 
 For a black hole of mass $M_{BH}$ (also measured in Planck scale units) and the size $R$, this gap is at least by a factor  $\sim 1/M_{BH}^2$ more narrow than a quantum level-spacing in any ordinary system of the same size
 (for more details, see the counting in 
 \cite{NPortrait,quantumPhase, goldstone, mischa1}).   
 Secondly,  the number of qudits must scale as the black hole area. 
 Nothing certain is known about how a black hole manages to achieve this goal. 
 The key problem is the lack of non-perturbative techniques that would allow to perform computations beyond the semi-classical picture.  
 
 One legitimate approach for shedding light on mysterious properties of {\it incalculable}  systems is to look for analogous properties in systems  that are much easier to solve.  If this search is successful, it can enable us to understand the basic mechanism in terms of some universal phenomenon that takes place also in non-gravitational systems.   
 This strategy was first adopted in \cite{NPortrait, quantumPhase}, where
 a system of attractive bosons in a box of dimensionality $D \geq 1$ with periodic boundary conditions was studied.\footnote
 {A simple example of this sort,  which has been explicitly solved, is given by a gas of bosons with four-point interaction in $D=1$ (see \cite{kanamoto}).  It is the model that we use in appendix \ref{app:periodic} to demonstrate our $c$-number method.}
 It was noticed that the critical state of enhanced memory capacity in this simple system exhibits some similarities to certain universal scaling properties satisfied by analogous parameters for black holes. Therefore, it was hypothesized in \cite{NPortrait, quantumPhase} that 
 the emergence of gapless qudits responsible for nearly degenerate microstates of a black hole in its bare essence 
 is the same phenomenon as the appearance of gapless qudits around the quantum critical point in the system of attractive bosons.

 This point of view is further strengthened by the  
model of   \cite{areaLaw} (see \cite{neural2} for its neural network realization), which describes a non-relativistic bosonic   quantum field  
 living on a $D$-dimensional sphere and experiencing a momentum-dependent attractive self-interaction. 
 This model comes closest to imitating the black hole information properties.  Namely, it exhibits a 
one parameter family of critical states labeled 
by the occupation number $N$ of the lowest momentum mode.  In each state, a set of gapless modes emerges.  The remarkable thing is that the number of these gapless 
 modes scales as the area of a $D-1$-dimensional sphere. This means that the resulting microstate entropy  obeys an area law, similar to Bekenstein entropy  of a black hole.
 Therefore, the  emergent gapless modes represent 
 holographic degrees of freedom and the model gives an explicit 
 microscopic realization of the idea of holography, which is usually 
 considered to be an exclusive property of gravitational systems, such as black holes\cite{hologBH} or 
 AdS-spaces \cite{hologAdS}.  These findings give a strong motivation for further  studying the 
 proposed mechanism of emergence of gapless modes 
 at criticality in systems of bosons with "gravity-like" attractive interactions.

 Another motivation for our study is that the emergence of gapless modes could be a mechanism to enhance the memory capacity in neural networks. Indeed,  as suggested in \cite{neural1}, 
 the above discussed phenomenon of critical memory enhancement due to assisted gaplessness can establish a connection between the underlying mechanisms of the enhanced 
 memory  storage capacities in black holes and in quantum brain neural networks. 
 Furthermore, as shown in \cite{neural2}, an explicit mapping of a neural network onto a system of cold bosons can be achieved by 
 identifying the neural degrees of freedom with the different momentum modes of  bosons and simultaneously  identifying the excitatory synaptic connections between the neurons with the 
 couplings between the different momentum modes.

  Thus, the next logical step would be to start performing studies in experimental settings, i.e., to bring a gas of attractive cold bosons to the critical point and to measure various characteristics of the system
 predicted by theoretical studies.  Such experiments  have never been 
 performed previously. On the one hand, this would enable us to study the information processing properties of cold bosons, which are interesting in their own right. In particular, this opens up the possibility to create efficient storers of quantum information. On the other hand, the above connection can give an interesting prospect of simulating in table top quantum experiments the key mechanism of information storage in such seemingly-remote systems as black holes and quantum brain neural networks.
 
 As discussed above, one of the biggest mysteries in black hole physics is the origin of nearly-gapless qubits (or modes) that store quantum information. It has been 
hypothesized \cite{NPortrait, quantumPhase, nico, goldstone, mischa1, mischa3} that the general mechanism behind the emergence 
of such gapless modes can be understood as a many-body phenomenon of 
the type discussed in the current paper. Since 
in the present work we theoretically demonstrate the existence of 
assisted gaplessness already in a simple one-dimensional system,  this naturally opens up a prospect of possible experimental simulations. 
Such simulations could both verify the proposed phenomenon and check how far the similarity with black hole information processing goes. 

For example, a very concrete effect to check would be how the 
timescale of information storage near criticality scales with $N_0$, which plays the role of a macroscopic parameter analogous to the black hole mass.
Another  interesting question concerns the scrambling and a release of information: When an excitation is added to a system of enhanced memory capacity, how fast does it get entangled with the rest of the system and how long does it take to read it out afterwards?
Obviously,  the main focus is 
on understanding generic features of enhanced information storage and by no means on imitating intrinsically geometrical properties of black holes.

\subsection{Choice of Concrete Prototype System} 

In order to demonstrate the $c$-number method for finding critical states of enhanced memory storage, we shall apply it to a concrete prototype system. We consider a gas of cold bosons that experience a simple attractive contact interaction and are placed in a one-dimensional box with Dirichlet  boundary conditions. We truncate the tower of momentum modes to three and thus end up with a system of three interacting 
bosonic quantum modes with a specific Hamiltonian. 
Despite the expected simplicity of this 3-mode system,  we shall discover that it exhibits a rich variety of quantum phases. The nature of  quantum phase transitions is qualitatively different from the analogous system with periodic boundary conditions, which has been studied previously \cite{kanamoto, quantumPhase}  (see review in appendix \ref{app:periodic}).   

Since the three modes are bosonic, each of them can be in many different states labeled by its occupation number. 
Thus, from a quantum information point of view, they re\-present three qudits.  The attractive interaction translates as negative 
interaction energy between different modes.  Correspondingly, the system satisfies the conditions discussed above for the emergence of a nearly-gapless mode and for reaching the critical states of enhanced memory capacity. 
  Our goal is to identify such states with our analytic method and then confirm their existence by numerical analysis.

In the light of models of the type \cite{areaLaw}, which are easy to access analytically and which have a 
much closer connection with black hole entropy, the natural question that the reader may ask is why we are not focusing on those as opposed to the one-dimensional case with non-periodic boundary conditions. 
There are two reasons for this. The first one is presumed experimental simplicity. It should be easier to realize a simple contact interaction, as opposed to the momentum-dependent one considered in \cite{areaLaw}. Moreover, the prototype models studied so far only used periodic boundary conditions. For an experimental realization, however, it is important to determine how sensitive the phenomenon of emergence of gapless modes is to boundary conditions. In particular, non-periodic boundary conditions may also be easier to attain in an experimental setting.

Needless to say, we are aware of the extraordinary difficulties in performing such experiments. Therefore, what we present is not a concrete experimental proposal. In particular, we do not discuss any of the problems that arise due to an imperfect isolation from the environment.\footnote
{We could try to understand the disturbance effects due to the environment as a fluctuation $\delta N$ of the particle number. Since we expect it to grow slowly with the total number $N$ of bosons, it is clear that the relative disturbance $\delta N/N$ shrinks when we increase $N$. This suggests that choosing a large number of bosons, which is required in any case for decreasing the gap of the light modes, could also help to suppress disturbance effects from the environment.}
However, in the light of recent experimental progress (see \eg \cite{experimental} for experiments with cold atoms), the realization of a system that shares the key properties of our prototype model -- in particular the emergence of gapless modes -- seems a viable goal. We hope that the study of our prototype model can contribute to finding such a system.

The second reason for the choice of our prototype model is that the non-periodic and non-derivative case 
is harder to analyze analytically, and therefore it represents a better test of the $c$-number method. To put it shortly, we trade a simpler-solvable model  with a higher entropy
 for a harder-analyzable one with a smaller entropy
due to the idea that the latter model promises more experimental simplicity 
and tougher theoretical test of our method.  
The price we pay for this choice is that our system only produces a single gapless mode at the critical point. Of course, since 
we are in a one-dimensional system, the area-law strictly speaking is not well-defined.  
 Nevertheless, it suffices to illustrate the key qualitative point of assisted gaplessness in a simple setup with potential experimental prospects.

  \section{The $C$-Number Method} 

  \label{sec:analysis}
  
  \subsection{Procedure}
  We consider a set of $K+1$ bosonic quantum modes described by the creation and annihilation operators  
  $\hat{a}_k^{\dagger}$, $\hat{a}_k$  (where $k =0,1,\ldots,K$), which satisfy the standard canonical commutation relations \eqref{algebra}.  
  For later convenience, we introduce the notation
  	\begin{equation}
  	\hat{\vec{a}} = (\hat{a}_1, \ldots,\hat{a}_K)\,, \qquad \hat{\vec{a}}^\dagger = (\hat{a}_1^\dagger, \ldots,\hat{a}_K^\dagger) \,,
  	\end{equation}
  	where no distinction will be made between a vector and its transpose.
  	The reason for singling out one of the modes, in our notation $\hat{a}_0$, will become apparent shortly.
  We assume that the dynamics of the system is governed by a generic Hamiltonian,
  \begin{equation} \label{Ham}
  \hat{H} = \hat{H}(\hat{\vec{a}}^{\dagger}, \hat{\vec{a}}, \hat{a}_0^\dagger, \hat{a}_0) \,.
  \end{equation} 
 A priori, we do not have to put any restriction on it, i.e., we expect our method to work when  \eqref{Ham} depends on all possible normal-ordered interactions of the modes. But the concrete application of the $c$-number method will be sensitive to the symmetries of the Hamiltonian. The reason is that any symmetry also leads to a gapless transformation.\footnote
  	{The simplest example is the Hamiltonian of a non-interacting mode,
  		\begin{equation}
  		\hat{H} = \hat{a}^\dagger \hat{a} \,,
  		\end{equation}
  		which possesses a global $U(1)$-symmetry, $\hat{a} \rightarrow \ex^{i\varphi} \hat{a}$, due to particle number conservation. So the states $\ket{\Psi(\hat{a})}$ and $\ket{\Psi(\ex^{i\varphi}\hat{a})}$ have the same expectation values of the energy, but this is not connected to attractive interaction.}
  	However, we do not want to consider those but solely focus on the ones that arise due to a collective attractive interaction.
  
  For the sake of simplicity, we will not consider the case of generic symmetries, but focus on a special case of particular physical importance. We assume that the Hamiltonian only possesses one symmetry, namely a global $U(1)$-symmetry due to particle number conservation. So the generic Hamiltonian reads
  \begin{eqnarray} \label{HGeneral}
  \hat{H}  & = & \sum_{k=0}^K  E_k  \hat{a}_k^{\dagger} \hat{a}_k \,  + \,  
  \sum_{k,j,m,n=0}^K  \alpha^{(4)}_{kjmn}  \hat{a}_k^{\dagger}  \hat{a}_j^{\dagger} \hat{a}_m  \hat{a}_n \,   \\ \nonumber\
  & + & \, \sum_{k,j,m,n,o,p=0}^K  \alpha^{(6)}_{kjmnop}  \hat{a}_k^{\dagger}  \hat{a}_j^\dagger \hat{a}_m^\dagger\hat{a}_n  \hat{a}_o \hat{a}_p  \,  \\ \nonumber\
  & + & \, \sum_{k,j,m,n,o,p,q,r=0}^K  \alpha^{(8)}_{kjmnopqr}  \hat{a}_k^{\dagger}  \hat{a}_j^{\dagger} \hat{a}_m^{\dagger} \hat{a}_n^{\dagger}   \hat{a}_o \hat{a}_p \hat{a}_q \hat{a}_r\, + \,\ldots\,, \nonumber
  \end{eqnarray} 
  where $E_k,  \alpha^{(4)}_{kjmn}, \alpha^{(6)}_{kjmnop}, \alpha^{(8)}_{kjmnopqr}, \ldots$ are some parameters. We shall assume that the full Hamiltonian is bounded from below, but some interaction terms can be negative so that 
  the energy landscape is non-trivial.   
  
We are interested in the phenomenon of assisted gaplessness, \ie we would like to identify states around which  a high occupation of some modes assists other in becoming gapless. 
  As explained, the nearly-gapless modes will lead to a neighborhood in the Fock space where a large number of states fits within a narrow energy gap. This causes the enhanced memory capacity in which we are interested.
  We shall show that only two conditions suffice to form  such a neighborhood of high microstate density, generated by some emergent gapless mode(s). The first one is that the degrees of freedom are bosonic so that some of the modes can be highly occupied. The second one is that the interaction energy among the modes can assume negative 
  values.

  Finding such critical states requires a diagonalization of the Hamiltonian, which in general is computationally a very hard task. Our goal is to show that 
  under certain conditions, the diagonalization procedure
  can be 
  substituted by a much simpler approach of finding an extremum 
  of a $c$-number function. To this end, we perform the Bogoliubov approximation\cite{bogoliubov}, i.e., we replace the creation and annihilation operators by $c$-numbers,  
  \begin{subequations} \label{bogoliubovReplacement}
  	\begin{align}
  	\hat{\vec{a}} \rightarrow \vec{a}\,, &\qquad  \hat{\vec{a}}^{\dagger} \rightarrow \vec{a}^*\,, 	\label{replace}\\
  	\hat{a}_0 \rightarrow \sqrt{N-\sum_{k=1}^K |a_k|^2}\,, & \qquad 	\hat{a}_0^\dagger \rightarrow \sqrt{N-\sum_{k=1}^K |a_k|^2} \,,	\label{replace0}
  	\end{align} 
  \end{subequations}
  where $a_k$ are complex numbers and we introduced the abbreviation
  	\begin{equation}
  	\vec{a} = (a_1, \ldots, a_K)\,, \qquad  \vec{a}^* = (a_1^*, \ldots, a_K^*) \,.
  	\end{equation}
  	Note that we have replaced $K+1$ quantum modes by only $K$ complex variables. The reason is particle number conservation, as will become apparent in the proof of our method. Because of it, the sum of the moduli are fixed and moreover we have to fix a global phase.
  	Furthermore, note that particle number conservation as in \eqref{replace0} shows that the complex numbers scale as $a_i\sim \sqrt{N}$. In summary, we obtain the replacement
  \begin{equation} \label{hamiltonianReplacement}
  \hat{H}(\hat{\vec{a}}^{\dagger}, \hat{\vec{a}}, \hat{a}_0^\dagger, \hat{a}_0) \rightarrow H_{\text{bog}}(\vec{a}, \vec{a}^*) \,,
  \end{equation}
  where $H_{\text{bog}}(\vec{a}, \vec{a}^*)$ is an algebraic $c$-number function, which depends on $K$ complex variables. 
  
  We expect that the error in the Bogoliubov approximation scales as $1/N$. Thus, we can make it arbitrarily small in the double-scaling limit of large particle number:
  	\begin{equation}
  	N\rightarrow \infty\,, \qquad \alpha^{(i)} \rightarrow 0\,, \qquad  \text{with}\ \lambda^{(i)} \equiv \alpha^{(i)} N^{i/2-1} = \text{const.} \,,
  	\label{doubleScaling}
  	\end{equation}
  	where we suppressed the indices of the coupling constants.
  	The above limit corresponds to taking the individual interaction strengths to zero in such a way that the {\it collective } couplings $\lambda^{(i)}$ stay constant. Note that for the special case of 4-point interaction, we obtain the collective coupling $\lambda = \alpha N$. 	 
	 Throughout this paper, the limit $N\rightarrow \infty$ will refer to the double-scaling \eqref{doubleScaling}. In this limit,  the $c$-number method for finding gapless modes will be exact. For finite $N$, corrections appear that scale as a power of $1/N$.
  
 Before we can come to the main statement of this section, we introduce the notion of a {\it critical point} of the Bogoliubov Hamiltonian $H_{\text{bog}}$. It is defined as a value  $\vec{a}_\circ$ such that the first derivative vanishes,
  \begin{equation}\label{extremum}
  \frac{\partial H_{\text{bog}}}{\partial \vec{a}}\bigg|_{\vec{a} = \vec{a}_\circ} = 0 \,, 
  \end{equation} 
and moreover the determinant of the second derivative matrix is zero, 
  \begin{equation}\label{inflection}
  \det \mathcal{M}\Big|_{\vec{a} = \vec{a}_\circ} = 0 \,, \qquad \text{where} \ {\mathcal M} \equiv  
  \begin{pmatrix}
  {\mathcal B}^*   &   {\mathcal A}  \\
  {\mathcal A}^T    &    {\mathcal B}
  \end{pmatrix} \,.
  \end{equation}  
  Here the matrices ${\mathcal A}$ and  ${\mathcal B}$ denote 
  $ {\mathcal A}_{kj} \equiv {\partial^2 H_{\text{bog}} \over \partial a_k^* \partial a_j}$
  and ${\mathcal B}_{kj} \equiv  {\partial^2 H_{\text{bog}} \over \partial a_k \partial a_j}$, which implies $B^T = B$ and $A^\dagger = A$.
  So we deal with a {\it stationary inflection}  point of 
  the function $H_{\text{bog}}(\vec{a}, \vec{a}^*)$, i.e., a point at which the curvature vanishes in some directions. Our goal is to prove the following implication.
 
\textbf{Theorem:\ }  If the $c$-number function $H_{\text{bog}}$ possesses a critical point (in the above sense), this implies -- in the full quantum theory -- the existence of a state with emergent gapless modes, and correspondingly, with an enhanced microstate entropy.\footnote
  {\ Note that example \eqref{FH1} is a special case in which all $K$ modes become gapless. In general, conditions \eqref{extremum} and \eqref{inflection} only imply at least one gapless mode.}
   To put it shortly, any critical point
  is a point of an enhanced memory storage capacity.  
  \subsection{Proof}
  \label{ssec:proof}

  In order to see this, we will follow the known procedure for determining the spectrum of quantum fluctuations around a given state. Namely we consider the expectation value of the Hamiltonian in an arbitrary state, for which only the expectation value of the particle number is fixed:
  	\begin{equation}
  	N = \sum_{k=0}^K \braket{\hat{a}_k^\dagger \hat{a}_k} \,.
  	\end{equation}
  	In the following, expectation values will always refer to such a state. As explained, we moreover want to fix a global phase to exclude the gapless direction that arises due to the corresponding symmetry. Up to $1/N$-corrections, we therefore obtain
  	\begin{equation} \label{a0Replacement}
  	\braket{\hat{a}_0} \approx \braket{\hat{a}_0^\dagger} \approx \sqrt{N- \sum_{k=1}^K \braket{\hat{a}_k^\dagger \hat{a}_k}} \,.
  	\end{equation}
  	In this way, we can make particle number conservation manifest and obtain a Hamiltonian that only depends on $K$ modes.
  
  Next, we shift the remaining $K$ mode operators by the constants  corresponding to the above-discussed stationary inflection point of the $c$-number function $H_{\text{bog}}$,        
  \begin{equation}\label{operatorShift}
  \hat{\vec{a}} \rightarrow \vec{a}_\circ +  \hat{\vec{\alpha}}  \,,\qquad
  \hat{\vec{a}}^{\dagger}  \rightarrow \vec{a}_\circ^* +  \hat{\vec{\alpha}}^{\dagger} \,.  	
  \end{equation} 
  Obviously, the operators $\hat{\alpha}_k^{\dagger}$, $\hat{\alpha}_k$ satisfy commutation relations analogous to \eqref{algebra}. So the replacement \eqref{operatorShift} is always possible and exact, not only as an equation for the expectation values. But of course,
  the Hamiltonian is not diagonal in the new modes $\hat{\alpha}_k^{\dagger}$, $\hat{\alpha}_k$. 
  
  Now we want to expand the theory around a state in which the expectation values of the original $\hat{a}_k$-modes are given as $\braket{\hat{a}_k^\dagger \hat{a}_k} = |a_{\circ, k}|^2$. Thus, we write down the effective Hamiltonian in which we keep terms up to second order in the $\hat{\alpha}_k$-modes. Since $\vec{a}_\circ$    
  extremizes the $c$-number function $H_{\text{bog}}$, 
  terms linear in  $\hat{\alpha}_k$-modes are absent  from the Hamiltonian. Moreover, the $c$-numbers scale as $a_k \sim \sqrt{N}$ whereas the $\hat{\alpha}_k$ are independent of $N$. Thus, each additional factor of $\hat{\alpha}_k$ leads to a suppression by $1/\sqrt{N}$. So in the limit \eqref{doubleScaling} of large $N$, the second-order term dominates and the effective Hamiltonian takes the following form:
  \begin{equation}\label{HEFF}
  \braket{\hat{\mathcal H}} \, = \, H_0 +  \braket{\hat{\vec{\alpha}}^{\dagger}  {\mathcal A} \hat{\vec{\alpha}}} \, + \,  \frac{1}{2} \bigg (\braket{\hat{\vec{\alpha}} {\mathcal B} \hat{\vec{\alpha}}}\, + \,  \braket{\hat{\vec{\alpha}}^{\dagger} {\mathcal B}^* \hat{\vec{\alpha}}^{\dagger}} \bigg ) \,, 
  \end{equation}
  where the constant $H_0 \equiv  H_{\text{bog}}(\vec{a}_\circ, \vec{a}_\circ^*)$ denotes the value of the $c$-number function at the extremal point.
  Up to this irrelevant constant, we can rewrite the Hamiltonian in 
  block-matrix form:
  \begin{equation}\label{HEFF1}
  \braket{\hat{\mathcal H}} \, = \,
 \frac{1}{2} \braket{\begin{pmatrix}
  	\hat{\vec{\alpha}}^{\dagger}  & 
  	\hat{\vec{\alpha}}     
  	\end{pmatrix}
  	\begin{pmatrix}
  	{\mathcal B}^*   &   {\mathcal A}  \\
  	{\mathcal A}^T    &    {\mathcal B}
  	\end{pmatrix}
  	\begin{pmatrix}   
  	\hat{\vec{\alpha}}^{\dagger}    \\
  	\hat{\vec{\alpha}}     
  	\end{pmatrix}}
  +   {\rm const. }
  \end{equation}
  
  Now we can bring the Hamiltonian into a canonical diagonal form by 
  performing the following Bogoliubov transformation:  
  \begin{equation} \label{BMat}
  \begin{pmatrix}
  \hat{\vec{\alpha}}^{\dagger}    \\
  \hat{\vec{\alpha}}     
  \end{pmatrix}
  =
  \mathcal{T}
  \begin{pmatrix}   
  \hat{\vec{\beta}}^{\dagger}    \\
  \hat{\vec{\beta}}     
  \end{pmatrix}\,, \qquad \text{with} \ \mathcal{T} = \begin{pmatrix}
  V^*  & U   \\
  U^*  &  V 
  \end{pmatrix} \,,
  \end{equation} 
  or equivalently, 
  \begin{equation}\label{BTR}
  \hat{\alpha}_k = U_{kj}^*  \hat{\beta}_j^{\dagger} + V_{kj}\hat{\beta}_j \,,  
  \end{equation} 
  where $U$ and $V$ are the transformation matrices
  and $\hat{\beta}_j^{\dagger},\hat{\beta}_j$ are the 
  new modes that form a diagonal canonical basis.   
  The canonical commutation relations imply the conditions: 
  \begin{equation} \label{offDiagonalConditions}
  VV^{\dagger}  - U^*U^T =\idop\,, \qquad VU^{\dagger} - U^*V^T = 0 \,.
  \end{equation} 
  As always, we choose the matrices $U$ and $V$ such that off-diagonal terms, of the type $\hat{\beta}_j \hat{\beta}_k$ and $\hat{\beta}_j^\dagger \hat{\beta}_k^\dagger$, are absent from the Hamiltonian. This implies that $U,V$ satisfy 
  \begin{equation} \label{diagonalConditions} 
  U^{\dagger} {\mathcal A}^TV^* + 
  V^{\dagger} {\mathcal A}U^*  +  V^{\dagger} {\mathcal B}^*V^*
  +  U^{\dagger} {\mathcal B} U^* = 0 \,. 
  \end{equation} 
  In this way, we bring the Hamiltonian to the form
  \begin{equation} \label{generalBogoliubovHamiltonian}
  \braket{ \hat{\mathcal H}} \, = \, \braket{\hat{\beta}_k^{\dagger} {\mathcal E}_{kj}\hat{\beta}_j } \, + \, {\rm const.}\,,
  \end{equation} 
  where the matrix $ {\mathcal E}$ is given by
  \begin{equation} 
  {\mathcal E} \equiv U^{\dagger} {\mathcal A}^TU + 
  V^{\dagger} {\mathcal A}V  +  V^{\dagger} {\mathcal B}^*U
  +  U^{\dagger} {\mathcal B} V\, .
  \end{equation} 
  Note that the conditions \eqref{offDiagonalConditions} and \eqref{diagonalConditions} allow 
  the multiplication of $U$ and $V$ by an arbitrary unitary matrix. Therefore, without loss of generality, we can set the Hermitian matrix ${\mathcal E}$ to be diagonal. 
  
  Now, due to the fact that the modulus of the determinant of the matrix $\mathcal{T}$ is $1$, the condition $\det{\mathcal M} =0$ 
  is equivalent to the condition $\det {\mathcal E} =0$.\footnote
  {Following \cite{bogoliubovCalculation1}, we can infer the determinant of the matrix $\mathcal{T}$ from the relation 
  	\begin{equation} \label{determinantRelation}
  	\mathcal{T} \mathcal{J} \mathcal{T}^\dagger = \mathcal{J} \,,
  	\end{equation}
  	where $\mathcal{J} = \text{diag}(\idop,-\idop)$ and $\idop$ is a unit matrix of dimension $K$. The equality \eqref{determinantRelation} in turn is a consequence of the Bogoliubov conditions \eqref{offDiagonalConditions}.}
  Thus,  among the degrees of freedom described by 
  the operators $\hat{\beta}_k^\dagger,\hat{\beta}_k$, there exist gapless modes. 
  Moreover, the number of zero 
  eigenvalues of the two matrices is the same since multiplication by regular matrices does not change the dimension of the kernel of a matrix.
  So the number of gapless modes is given by the number of zero eigenvalues of the matrix  ${\mathcal M}$, i.e., by the number of independent flat directions at the critical point of the $c$-number function $H_{\text{bog}}$.
  
  This conclusion is exact in the limit \eqref{doubleScaling} of infinite $N$. In this case, the gap collapses to zero and the different quantum states that correspond to the different occupation numbers of the gapless modes become exactly degenerate. So the system can store an unlimited amount of information within an arbitrarily small energy gap. Note that the fact that we only make a statement about the expectation value of the Bogoliubov Hamiltonian in \eqref{generalBogoliubovHamiltonian} is not a restriction since it suffices for us to find states with degenerate expectation values of the energy.
  
  For finite $N$, corrections appear which scale as a power of $1/N$. They come from higher-order terms in the effective Hamiltonian \eqref{HEFF} and from corrections to relation \eqref{a0Replacement}. So in this case, the modes will only be nearly-gapless, with a gap that scales as a power of $1/N$. Also the critical value $\vec{a}_\circ$ will receive $1/N$-corrections. However, one can make all these corrections arbitrarily small if one chooses $N$ large enough. So also for finite $N$, the information stored in the various states of the
  	$\hat{\vec{\beta}}$-modes is energy cost-efficient. In summary, we conclude that the 
  critical point of the $c$-number function $H_{\text{bog}}$ corresponds to 
  the appearance of nearly-gapless modes in the full quantum theory.
  Each nearly-gapless mode corresponds to a zero eigenvalue of the  
  second derivative matrix ${\mathcal M}$.    
  
  We remark that our $c$-number method is conceptually similar to the study of the Gross-Pitaevskii equation \cite{grossPitaevskii}, which corresponds to working in position space and expanding the field operator $\hat{\psi}$ around its classical value: $\hat{\psi} = \psi_{\text{cl}} + \delta \hat{\psi}$. In this approach, one can identify gapless modes by studying the spectrum of quantum fluctuations $\delta \hat{\psi}$. Our $c$-number method can be viewed as momentum space analogue of this technique. Namely, we first go to momentum space by expanding $\hat{\psi}$ in mode operators $\hat{a}$. Then we proceed analogously to the Gross-Pitaevskii method by expanding mode operators around their classical values: $\hat{\vec{a}} = \vec{a}_\circ + \hat{\vec{\alpha}}$.

  \subsubsection*{Coherent State Basis}   
  
  Finally, we note that an alternative proof of the enhanced memory capacity around the stationary inflection point of $H_{\text{bog}}$ consists of moving to the basis of coherent states, as opposed to number eigenstates. 
  We recall that  coherent states  $\ket{\vec{a}}$ 
  are the eigenstates of the destruction operators, i.e., for all modes we have $\hat{a}_k
  \ket{\vec{a}} = a_k \ket{\vec{a}}$, where 
  $a_k$ is a complex eigenvalue. Obviously, coherent states satisfy
  $|a_k|^2 = \bra{\vec{a}} \hat{n}_k \ket{\vec{a}}$. 
  It is clear that taking an expectation value of the Hamiltonian 
  (\ref{HGeneral})  over a coherent state $\ket{\vec{a}}$ simply amounts to the Bogoliubov approximation \eqref{bogoliubovReplacement}, i.e.,
  to replacing the operators by $c$-numbers, $\hat{a}_k \rightarrow a_k$. Therefore, we have the relation 
  \begin{equation} \label{coherentStateExpectation}
  \bra{\vec{a}} \hat{H} \ket{\vec{a}} = H_{\text{bog}} \,.
  \end{equation}
  This means that coherent states explicitly realize the replacement \eqref{hamiltonianReplacement}.
  	Since this procedure is exact also for finite $N$, it gives immediate meaning to the Bogoliubov Hamiltonian from the perspective of the full quantum system.
  
  In particular, this construction is relevant when the Bogoliubov Hamiltonian possesses a stationary inflection point $\vec{a}_\circ$. If in this case the eigenvector with vanishing eigenvalue is given by $\vec{\delta a}$, then we can consider the state $\ket{\vec{a}_\circ + \epsilon\vec{\delta a}}$. For small values of $\epsilon$, it fulfills
  	\begin{equation}
  	\bra{\vec{a}_\circ + \epsilon\vec{\delta a}} \hat{H} \ket{\vec{a}_\circ + \epsilon\vec{\delta a}} = \bra{\vec{a}_\circ} \hat{H} \ket{\vec{a}_\circ} \,.
  	\label{degenerateStates}
  	\end{equation}
  	Thus, we have obtained a family  $\ket{\vec{a}_\circ + \epsilon\vec{\delta a}}$ of quantum states with nearly degenerate expectation value of the energy. Information stored in them therefore occupies a narrow gap.

  For quantifying the information storage capacity, we must take into account that coherent states do not form a orthonormal basis and that only coherent states with large enough differences
  in $a_k$ are nearly orthogonal. Indeed,  the scalar product of two coherent states  $\ket{\vec{a}}$ and $\ket{\vec{a}'}$ is 
  \begin{equation} \label{product}
  | \braket{\vec{a}| \vec{a}'} |^2 = \ex^{-\sum_k |a_k - a_k'|^2} \,.
  \end{equation} 
  Because of this, although the coherent state parameter 
  can take continuous values, only sufficiently distant states, which satisfy 
  \begin{equation}\label{distance} 
  \sum_k |a_k - a_k'|^2 \gg 1 \,,
  \end{equation} 
  contribute into the memory-capacity count.  
  Due to this,  the information storage capacity in the coherent state basis is the same as 
  in the basis of number eigenstates of the Bogoliubov modes.\footnote
  {Relation \eqref{distance} implies that coherent states can be counted as different as soon as $\left|n_k - n_k'\right| \gg \sqrt{n_k}$, where $n_k = |a_k|^2$ and $n_k' = |a_k'|^2$. So there are on the order of $\sqrt{n_k}$ different possible expectation values of the particle number. In addition, however, there is the freedom of choosing a phase $\varphi_k$. Taking into account the uncertainty, $\Delta n_k \Delta \varphi_k \gtrsim 1$, this gives $\sqrt{n_k}$ different phases for each modulus $n_k$. In sum, this gives $n_k$ different states, the same result as in the basis of number eigenstates.}
  However, the usefulness of the  coherent state basis lies in the ability of taking a smooth classical limit. This is convenient for the generalization of the enhanced memory  storage phenomenon to classical systems, such as e.g., classical neural networks \cite{neural1}.    
  
  For an exemplary step-by-step application of the $c$-number method we refer the reader to appendix \ref{app:periodic}. There we use it to study the periodic analogue of our prototype model, \ie the attractive one-dimensional Bose gas with periodic boundary conditions. 
  For this system, a complete analytic treatment is possible. This means that on the one hand, all equations resulting from the $c$-number method  can be solved easily and on the other hand, the Bogoliubov transformation can be carried out explicitly. It is therefore a good starting point to both familiarize oneself with the method and to check its validity on a concrete example.

\section{Prototype Model: 3-Mode System}
\label{sec:3Mode}
\subsection{Introduction of Bose Gas with Dirichlet Boundary Conditions}
 \label{ssec:basicHamiltonian}
 We proceed to apply the $c$-number method to our prototype model, the one-dimensional Bose gas in a box.
 Its Hamiltonian is given by
 \begin{equation}
 \hat{H} =  \limitint_{0}^{L} \ldiff{z}   \Big [ \frac{\hbar^2}{2m} \partial_z \hat{\psi}^\dagger \partial_z \hat{\psi}
 - \frac{\hbar^2}{2m}\frac{\pi^2 \alpha }{L}\hat{\psi}^\dagger
 \hat{\psi}^\dagger \hat{\psi} \hat{\psi} \Big ]\,,
 \label{periodicHamiltonian}
 \end{equation}
 where $\alpha$ is a dimensionless, positive coupling constant describing the attractive four-point interaction of the atoms and $L$ is the size of the system. We impose Dirichlet boundary conditions so that the free eigenfunctions read
 \begin{equation}
 \hat{\psi} = \sqrt{\frac{2}{L}}\sum_{k=1}^\infty \hat{a}_k \sin \left(\frac{k \pi z}{L}\right). 
 \label{eq:unperiodicExpansion}
 \end{equation}
 Going to momentum space, we then obtain 
\begin{align*}
\centering
\hat{H}^{\text{full}} = & \frac{4\pi^2\hbar^2}{2m L^2} \bigg[\sum_{k=1}^\infty \frac{ k^2}{4} \hat{a}_k^\dagger \hat{a}_k
    -  \frac{\alpha }{8} \sum_{k,l,m =1}^\infty
\big [(\hat{a}_k^\dagger \hat{a}_l^\dagger \hat{a}_m \hat{a}_{k+l-m}+
2 \hat{a}_k^\dagger \hat{a}_l^\dagger \hat{a}_m \hat{a}_{k-l+m})\\ &- 2
(\hat{a}_{l+m+k}^\dagger \hat{a}_l^\dagger \hat{a}_{m} \hat{a}_{k} +
\hat{a}_{k}^\dagger \hat{a}_l^\dagger \hat{a}_{m} \hat{a}_{k+l+m}) \big]\bigg] \,.
\numberthis \label{fullHamiltonian}
\end{align*}
 Both analytically and numerically, however, it is difficult to obtain explicit solutions of the full Hamiltonian \eqref{fullHamiltonian}. Therefore, we will truncate the system to the lowest three modes, $k\leq 3$. This is the smallest number of modes for which the non-periodic system behaves qualitatively differently from its analogue with periodic boundary conditions. Explicitly, we obtain after truncation:
 \begin{align*} 
 \hat{H} &= \frac{1}{4} \sum_{k=1}^3 k^2 \hat{a}_k^\dagger \hat{a}_k 
 - \frac{\alpha}{8} \bigg [ 3 \hat{a}_1^{\dagger\, 2}\hat{a}_1^2 
 +8 \hat{a}_1^{\dagger } \hat{a}_2^{\dagger }\hat{a}_1 \hat{a}_2 
 +2 \hat{a}_1^{\dagger \, 2}\hat{a}_2^2 
 +2 \hat{a}_2^{\dagger\, 2}\hat{a}_1^2 \\
 &+8 \hat{a}_1^{\dagger }\hat{a}_3^{\dagger }\hat{a}_1 \hat{a}_3 
 +2 \hat{a}_1^{\dagger \, 2}\hat{a}_3^2
 +2 \hat{a}_3^{\dagger \, 2}\hat{a}_1^2
 -2 \hat{a}_1^{\dagger \, 2}\hat{a}_1\hat{a}_3 
 -2 \hat{a}_1^{\dagger } \hat{a}_3^{\dagger }\hat{a}_1^2
 \\
 &+4\hat{a}_1^{\dagger } \hat{a}_2^{\dagger }\hat{a}_2 \hat{a}_3
 +4 \hat{a}_2^{\dagger } \hat{a}_3^{\dagger }\hat{a}_1 \hat{a}_2
 +2 \hat{a}_1^{\dagger } \hat{a}_3^{\dagger }\hat{a}_2^2
 +2 \hat{a}_2^{\dagger \, 2}\hat{a}_1 \hat{a}_3
 \\
 &+3 \hat{a}_2^{\dagger \, 2}\hat{a}_2^2
 +8 \hat{a}_2^{\dagger }\hat{a}_3^{\dagger }\hat{a}_2 \hat{a}_3
 +2 \hat{a}_2^{\dagger \, 2}\hat{a}_3^2
 +2 \hat{a}_3^{\dagger \, 2}\hat{a}_2^2
 +3 \hat{a}_3^{\dagger \, 2}\hat{a}_3^2 \bigg ]. \numberthis 	\label{truncatedHamiltonian}
 \end{align*} 
 This Hamiltonian defines the prototype system that we shall study in the following. For convenience, we set $L = 2\pi$ and $\hbar = 2m = 1$ from now on. Our subsequent task is to understand  the phase portrait of the Hamiltonian \eqref{truncatedHamiltonian} with the aim of  identifying an emergent gapless mode that leads to enhanced entropy states with long decoherence time and large information storage capacity.

To prepare the application of our analytic method, we now perform the Bogoliubov approximation. Clearly, since the conditions \eqref{extremum} and \eqref{inflection}, which define critical points of the Bogoliubov Hamiltonian, allow for a reparametrization of the complex variables contained in $\vec{a}$ and $\vec{a}^*$, we can use a different parametrization defined by\footnote
{The equivalence of the vanishing of the second derivatives in $\vec{a}$ and $x$ is non-trivial and only holds if there are no unoccupied modes, $a_k\neq 0$. Schematically, the reason is that $\frac{\partial H_{\text{bog}}}{\partial a} = a \frac{\partial H_{\text{bog}}}{\partial x}$ and therefore $\frac{\partial^2 H_{\text{bog}}}{\partial^2 a} = a^2 \frac{\partial^2 H_{\text{bog}}}{\partial^2 x} + \frac{\partial H_{\text{bog}}}{\partial x}$.}
 \begin{equation} 
 \hat{a}_1 \rightarrow \sqrt{N(1-x)}\cos(\theta) \,, \qquad
 \hat{a}_2 \rightarrow \sqrt{Nx}\ex^{i\Delta_2} \,, \qquad
 \hat{a}_3 \rightarrow \sqrt{N(1-x)}\sin(\theta)\ex^{i\Delta_3} \,.
 \label{macroscopicReplacements}
 \end{equation}
As it should be, this substitution already incorporates particle number conservation, i.e, we replace three modes by only two complex numbers, or equivalently two moduli and two phases.
Here $0\leq x \leq 1$ is the relative occupation of the 2-mode and $0\leq \theta \leq \pi/2$ characterizes how the remaining atoms are distributed among the 1- and 3-mode. Moreover, $\Delta_2$ and $\Delta_3$ are relative phases. The Bogoliubov Hamiltonian, which we obtain after plugging in the replacements \eqref{macroscopicReplacements} in the Hamiltonian \eqref{truncatedHamiltonian}, reads: 
	\begin{align*}
\frac{H_{\text{bog}}}{N} & =  \frac{1}{4} \left(1 + 3x +8(1-x)\sin^2(\theta)\right)  - \frac{\lambda}{8} \Big[ \sin^2(2\theta)(1-x)^2\big(\frac{1}{2} +  \cos(2\Delta_3)\big)
  \\
&+ 3 + 2x -2x^2  + 4 x(1-x) \Big(\cos(2\Delta_2)\cos^2(\theta) + \cos(2\Delta_2-2\Delta_3) \sin^2(\theta) \Big)\\
&+ 2\sin(2\theta)(1-x)\left(  x  \cos(2\Delta_2- \Delta_3) + \cos(\Delta_3)\Big( 2x - (1-x)\cos^2(\theta) \Big) \right)
\Big] \,. \numberthis \label{macroscopicHamiltonian}
\end{align*}
   
\subsection{Analysis of the Ground State}
\label{ssec:criticalGroundState}

\begin{figure}
	\begin{center}
		\includegraphics[width=0.53\textwidth]{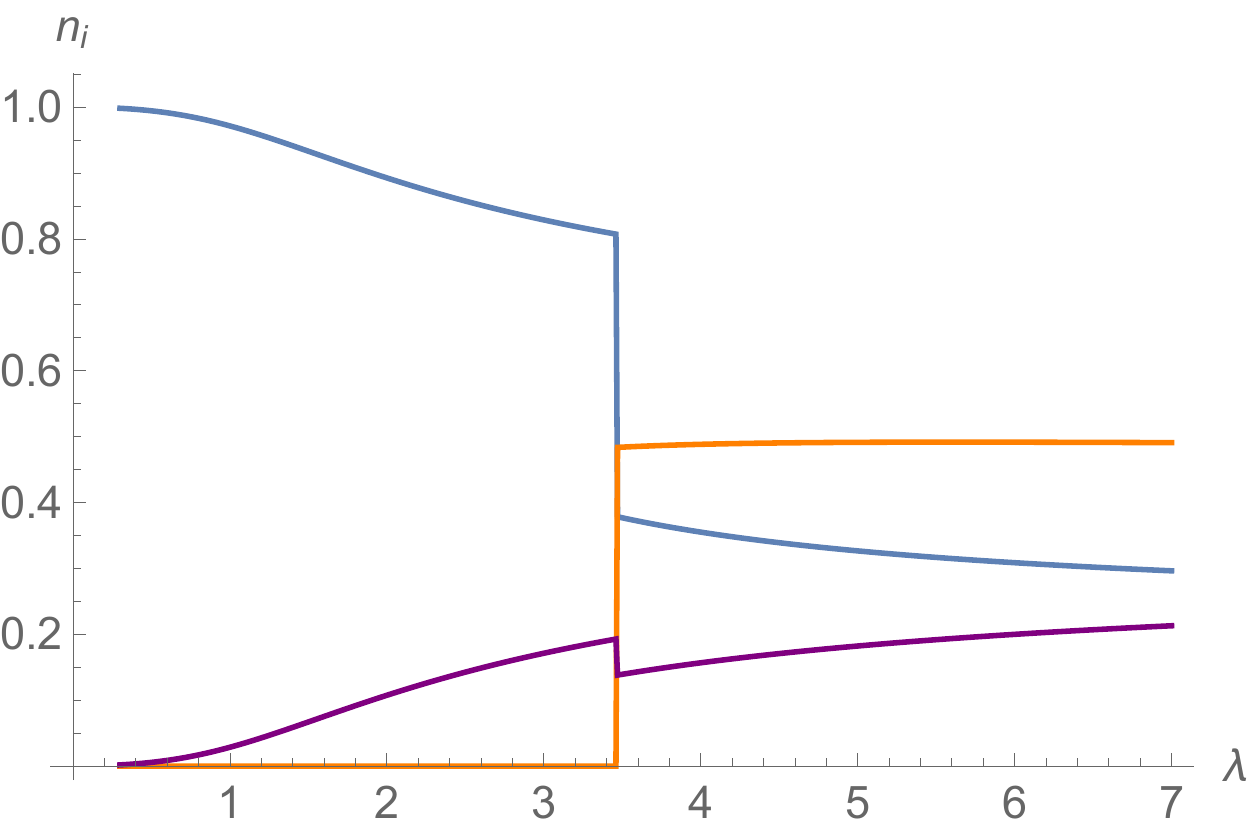}
		\caption{Relative occupation numbers of the ground state  of the Bogoliubov Hamiltonian as functions of $\lambda$. The 1-mode is displayed in blue, the 2-mode in orange and the 3-mode in purple. There is a discontinuous change in the occupation numbers at $\lambda_{gs} \approx 3.5$.}
		\label{fig:groundStateOccupation}
	\end{center}
\end{figure}

As a preparatory exercise, we analyze the ground state of the Hamiltonian \eqref{truncatedHamiltonian}. We can do so by finding the global minimum of the Bogoliubov Hamiltonian \eqref{macroscopicHamiltonian}. It is evident that the choices $\Delta_2=0$ as well as $\Delta_3=0$ or $\Delta_3=\pi$ are preferred since they minimize each term separately.\footnote
{This follow from the last line of the Hamiltonian \eqref{macroscopicHamiltonian}. For $3n_2>n_1$, $\Delta_3=0$ is preferred and otherwise $\Delta_3=\pi$.}
It is straightforward to minimize the energy with respect to $\Delta_3$ and the remaining two continuous parameters $x$ and $\theta$  numerically.\footnote
{All numerical computations in this work are performed with the help of Mathematica \cite{mathematica}. }
The resulting occupation numbers of the ground state as functions of the collective coupling $\lambda$ are displayed in \fig \ref{fig:groundStateOccupation}. We observe that the occupation numbers change discontinuously at the critical point $\lambda_{gs} \approx 3.5$, where the subscript $gs$ stands for ground state.

\begin{figure}
	\begin{center}
	\includegraphics[width=0.6\textwidth]{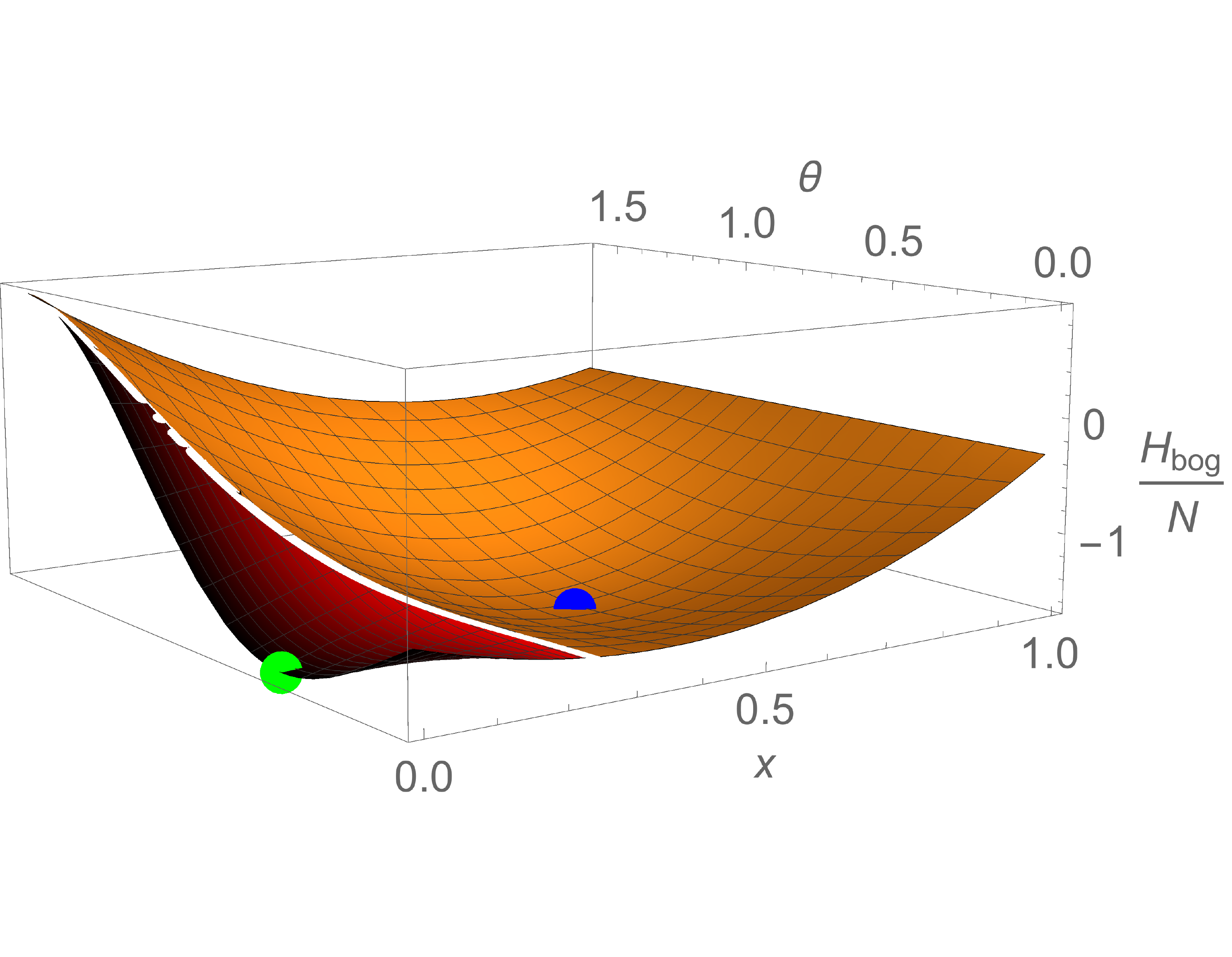}
		\caption{Bogoliubov energy (rescaled by the inverse particle number) for $\lambda=3.5$ as a function of $x$ and $\theta$. 
			 The red surface is the region where $\Delta_3=\pi$ minimizes the energy and in the orange surface, $\Delta_3=0$ is preferred. We observe two disconnected, degenerate minima, one for $x=0$ (green point) and one for $x\neq0$ (blue point).}

		\label{fig:HOfXAndTheta}
	\end{center}
\end{figure}

\begin{figure}
	\begin{center}	
		\begin{subfigure}{0.49\textwidth}
			\centering
			\includegraphics[scale=0.55]{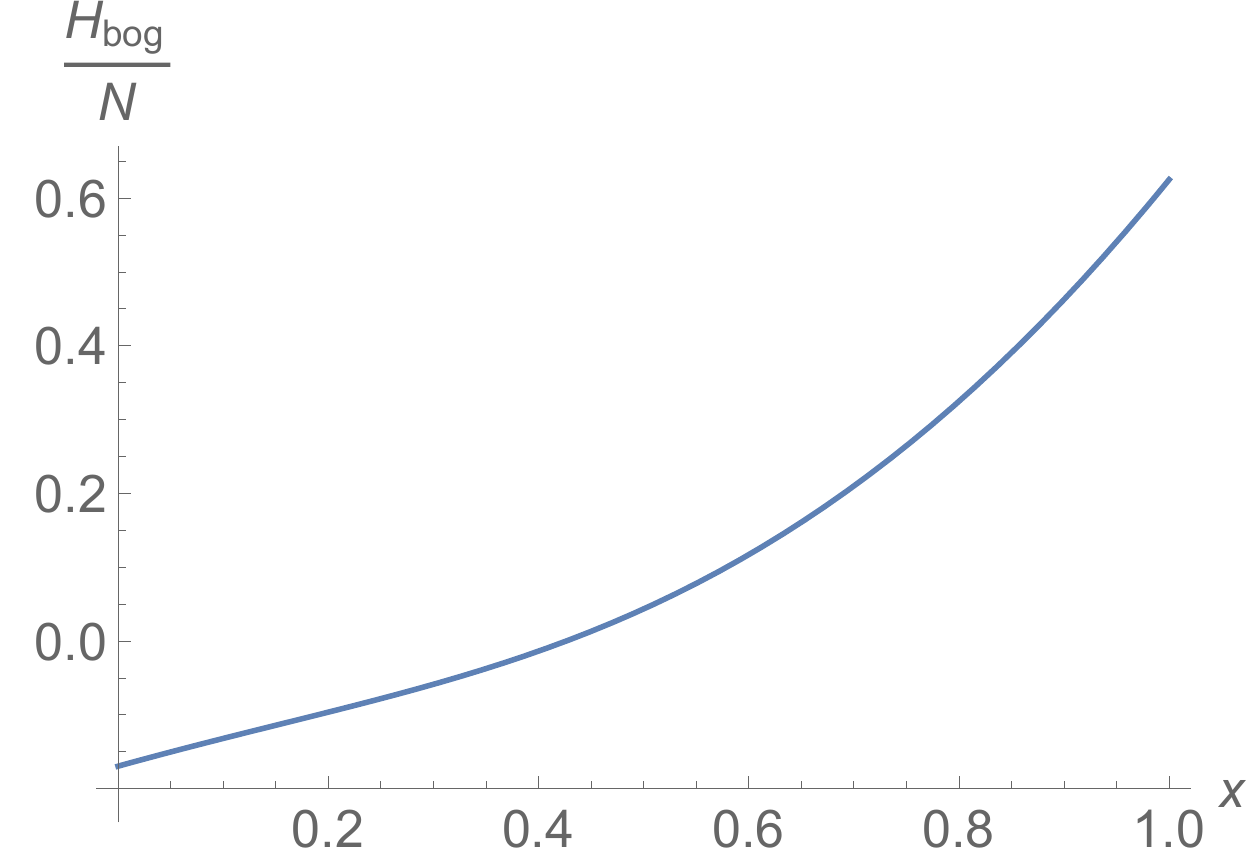}
			\caption{$\lambda = 1$}
			\label{fig:HOfXa}
		\end{subfigure}
		\begin{subfigure}{0.49\textwidth}
			\centering
			\includegraphics[scale=0.55]{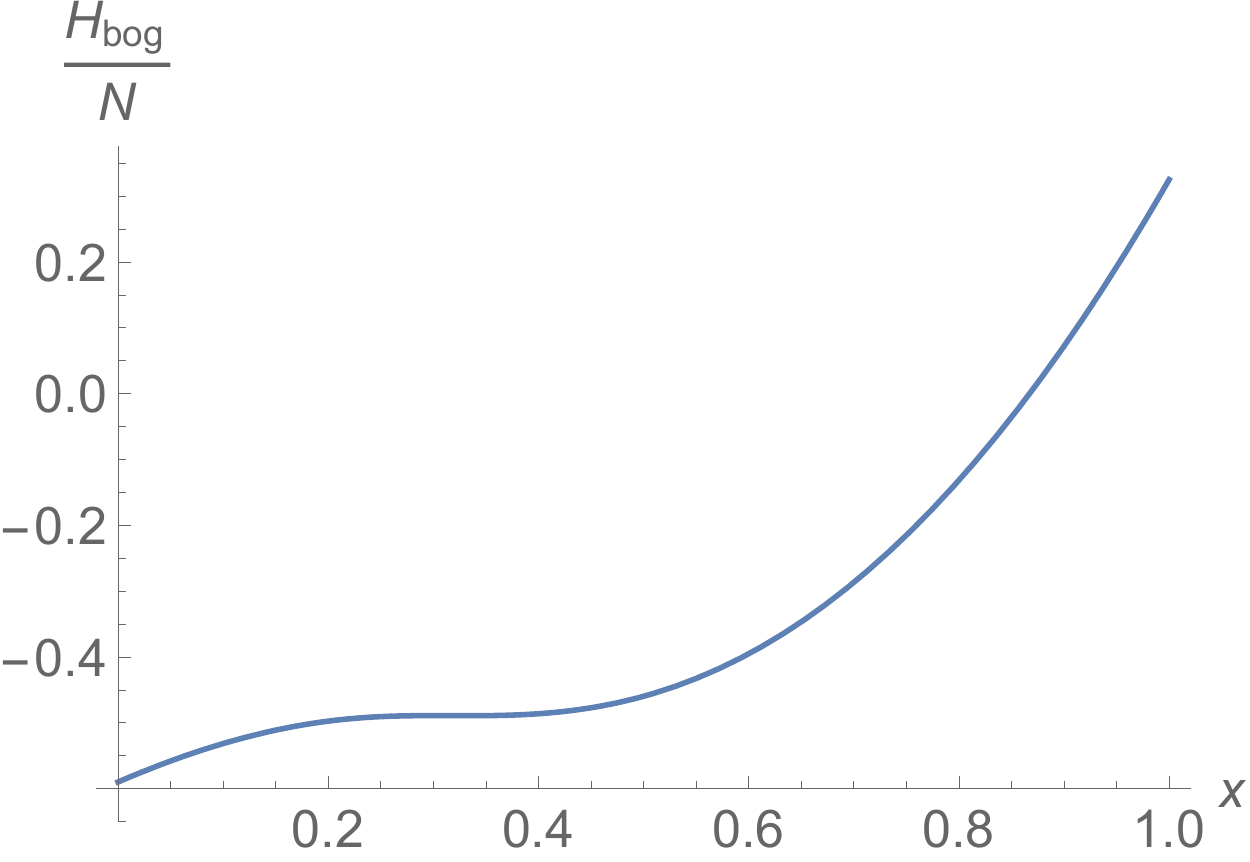}
			\caption{$\lambda = 1.8$}
			\label{fig:HOfXb}
		\end{subfigure}
		\begin{subfigure}{0.49\textwidth}
			\centering
			\includegraphics[scale=0.55]{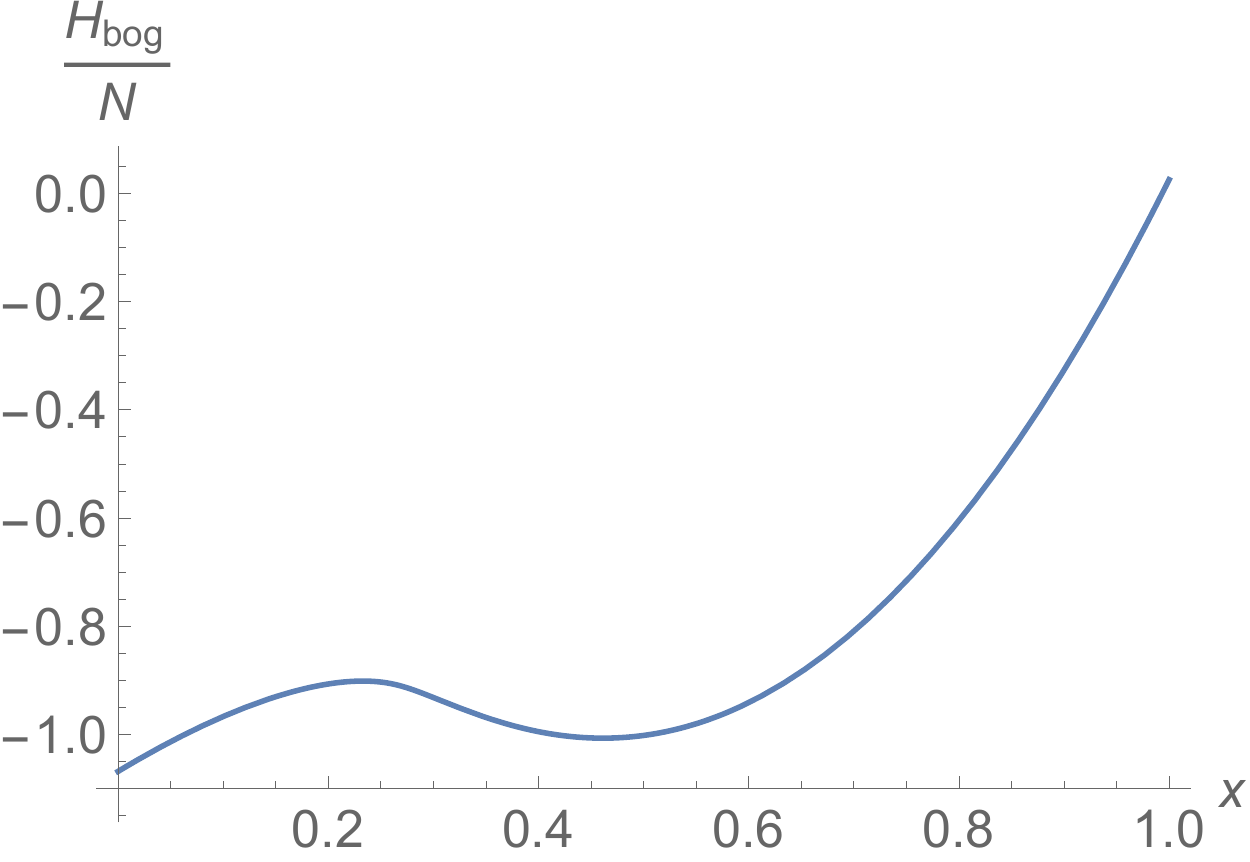}
			\caption{$\lambda = 2.6$}
			\label{fig:HOfXc}
		\end{subfigure}
		\begin{subfigure}{0.49\textwidth}
			\centering
			\includegraphics[scale=0.55]{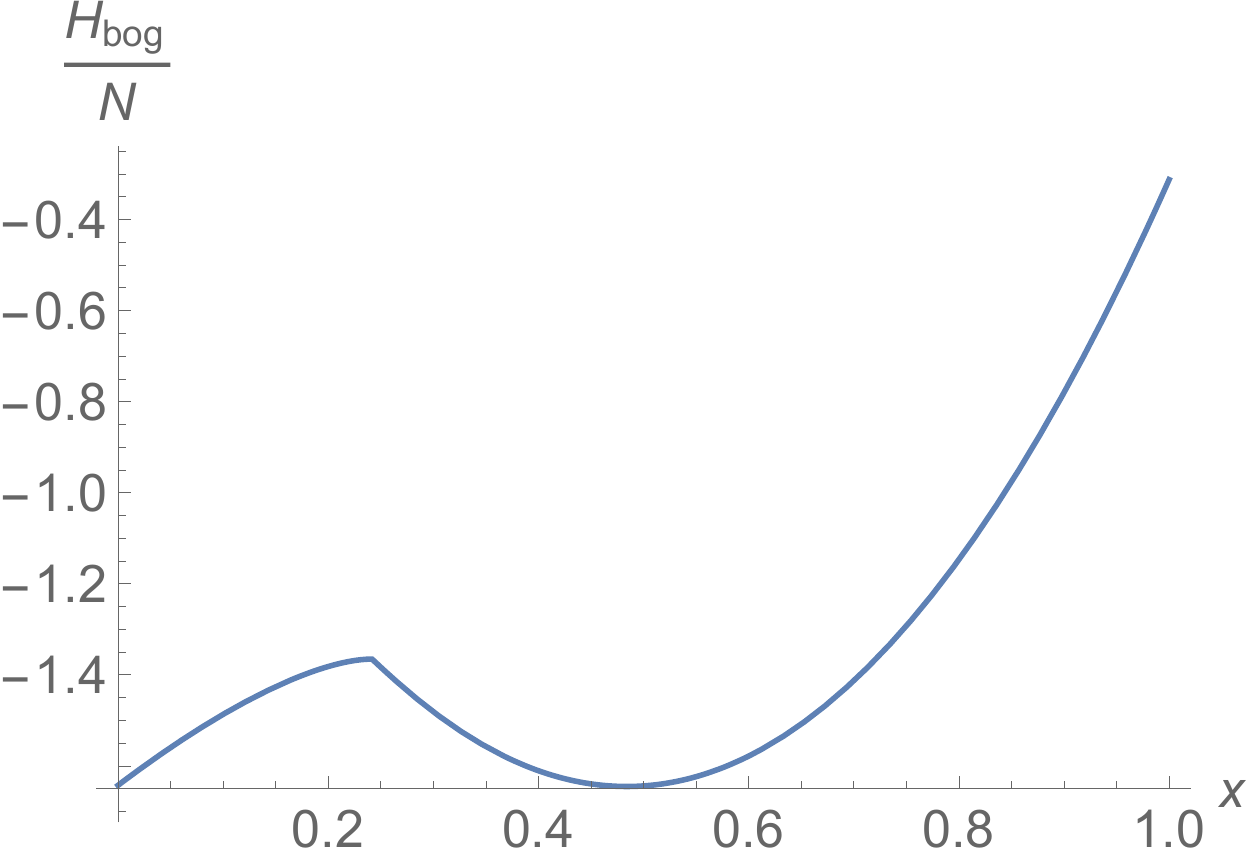}
			\caption{$\lambda = 3.5$}
			\label{fig:HOfXd}
		\end{subfigure}
		\begin{subfigure}{0.49\textwidth}
			\centering
			\includegraphics[scale=0.55]{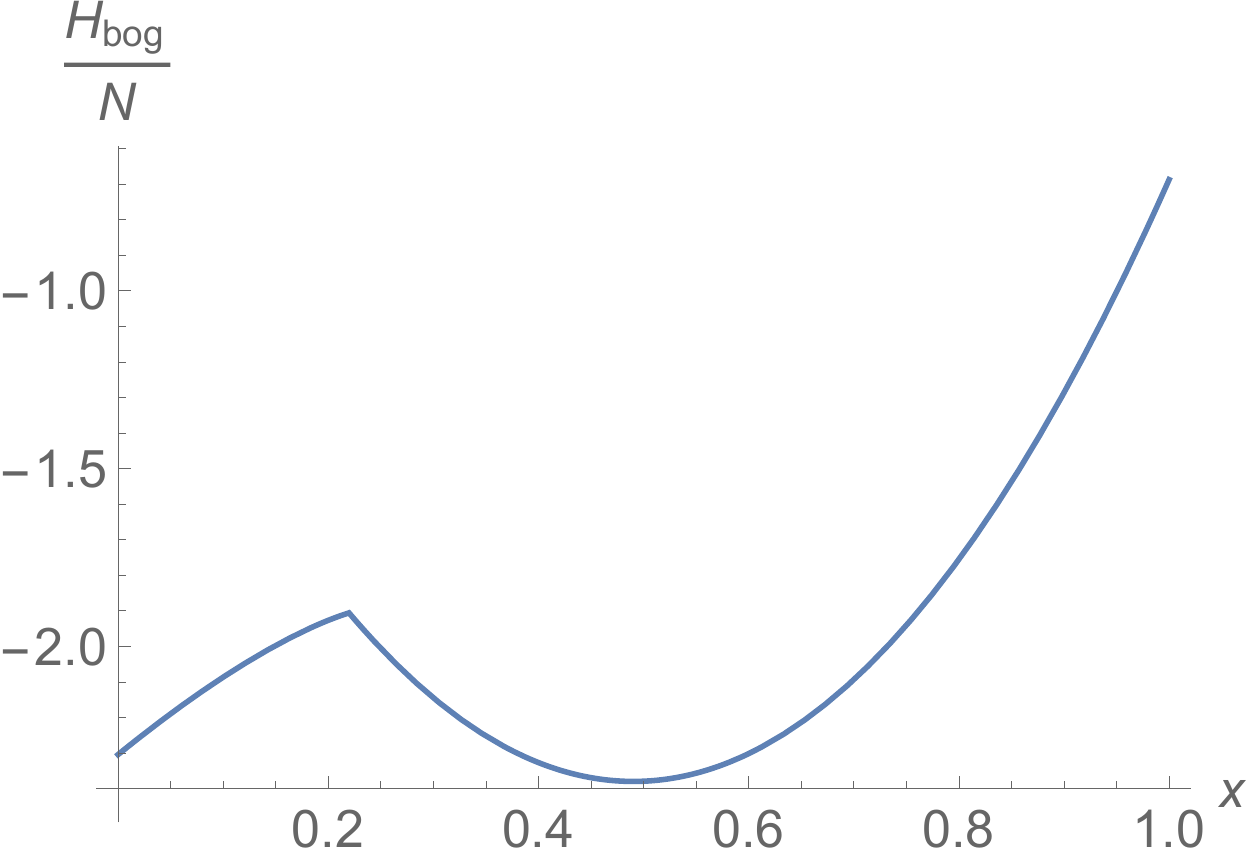}
			\caption{$\lambda = 4.5$}
			\label{fig:HOfXe}
		\end{subfigure}
		\caption{ Minimal value of the Bogoliubov Hamiltonian (rescaled by the inverse particle number) subject to the constraint that the relative occupation of the 2-mode is $x$. 
			 At $\lambda_{lm} \approx 1.8$, a stationary inflection point signals the appearance of a second minimum and at $\lambda_{gs} \approx 3.5$, this second minimum becomes energetically favorable.}
		\label{fig:HOfX}
	\end{center}
\end{figure}

 In order to understand this behavior better, we plot the Bogoliubov Hamiltonian as a function of $x$ and $\theta$ for the critical value $\lambda = 3.5$ in \fig \ref{fig:HOfXAndTheta}. Since we observe two disconnected, degenerate minima, we can explain the discontinuous change of the occupation numbers as transition between the two minima.
  To analyze how the second minimum develops, we marginalize over $\theta$ and $\Delta_3$, i.e., we only fix $x$ and minimize the energy with respect to the remaining parameters $\theta$ and $\Delta_3$. \fig \ref{fig:HOfX} shows the result for different values of $\lambda$. We conclude that a local minimum exists at $x=0$ for all values of $\lambda$ and that another local minimum at $x =x_{\text{min}}(\lambda) \neq 0$ starts to exist for $\lambda > \lambda_{lm}$,
  where
  \begin{equation}
  	 \lambda_{lm} \approx 1.8 \,.
  \end{equation}
 Here the subscript $lm$ stands for light mode since we observe that the critical point $\lambda_{lm}$ corresponds to a stationary inflection point and will therefore be crucial for our discussion of gapless modes in the next section. With regard to the ground state, we can conclude for now that $\lambda_{gs}$ corresponds to the point where the second minimum becomes energetically favorable. 

Thus, we expect from the analytic analysis that the ground state changes discontinuously at the critical point $\lambda_{gs}\approx 3.5$, \ie that there is a first-order phase transition. We can check that this indeed happens in the quantum Hamiltonian for finite $N$. To this end, we diagonalize it numerically to find the true ground state. When we plot the expectation values of the occupation numbers of the ground state as functions of $\lambda$, the result is indistinguishable from \fig \ref{fig:groundStateOccupation} above already for $N\gtrsim 100$. We therefore confirm that there is a critical point at $\lambda_{gs}$, at which the ground state of the system changes discontinuously.

This represents a marked difference to the periodic system, where the ground state changes continuously, \ie a second-order phase transition takes place (see appendix \ref{app:periodic}). Because of the continuity of the transition, higher modes only get occupied slowly in that case so that one can describe the full system solely in terms of the lowest three modes. This makes it easy to obtain numerical results for the periodic system. As we have seen, however, the occupation numbers change discontinuously for Dirichlet boundary conditions. Therefore, the truncation to three modes is no longer justified already near the critical point. For this reason, the full system \eqref{fullHamiltonian} does not necessarily need to exhibit the behavior which we observe for the truncated Hamiltonian \eqref{truncatedHamiltonian}.

\section{Critical Point with Gapless Mode}
\label{sec:criticalLightModes}

\subsection{Application of $C$-Number Method}
In this section, we perform a detailed analysis of the point $\lambda = \lambda_{lm}$.  Our goal is to show that it features a light mode and correspondingly an increased density of states,  i.e., that the phenomenon of assisted gaplessness takes place at this point. First, we will do so using the analytic $c$-number method developed in section \ref{sec:analysis}. As explained there, it allows us to forgo the involved analysis of the full spectrum. Instead, we are only faced with the much simpler task of showing that the Bogoliubov Hamiltonian, which only depends on two complex variables, possesses a stationary inflection point.

\begin{figure}
\centering 
\begin{subfigure}{0.4\textwidth}
\includegraphics[width=\textwidth]{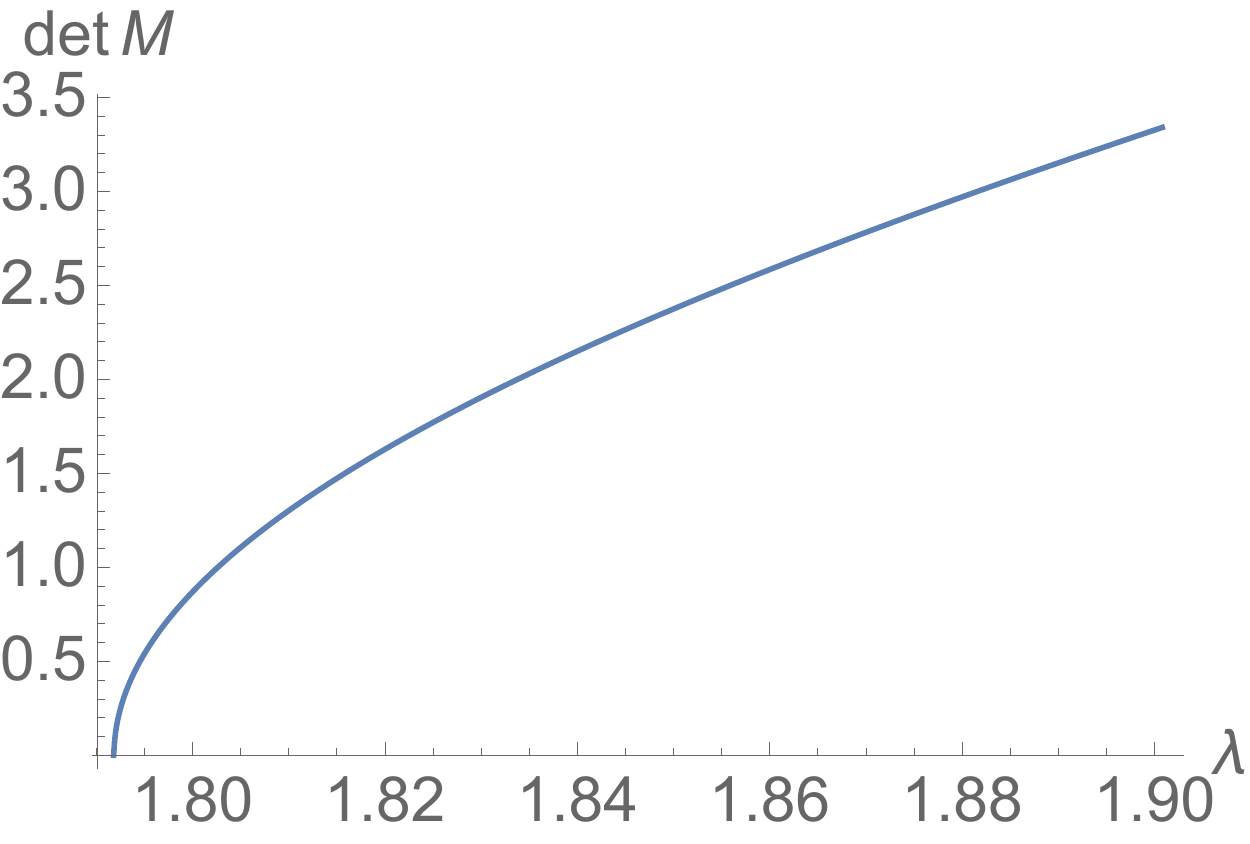}
\caption{Determinant of the second derivative \eqref{macroscopicHamiltonianSecondOrder} of the Bogoliubov Hamiltonian.} 
\label{fig:macroscopicFluctuations}
\end{subfigure}
\hspace{0.1\textwidth}
\begin{subfigure}{0.4\textwidth}
\includegraphics[width=\textwidth]{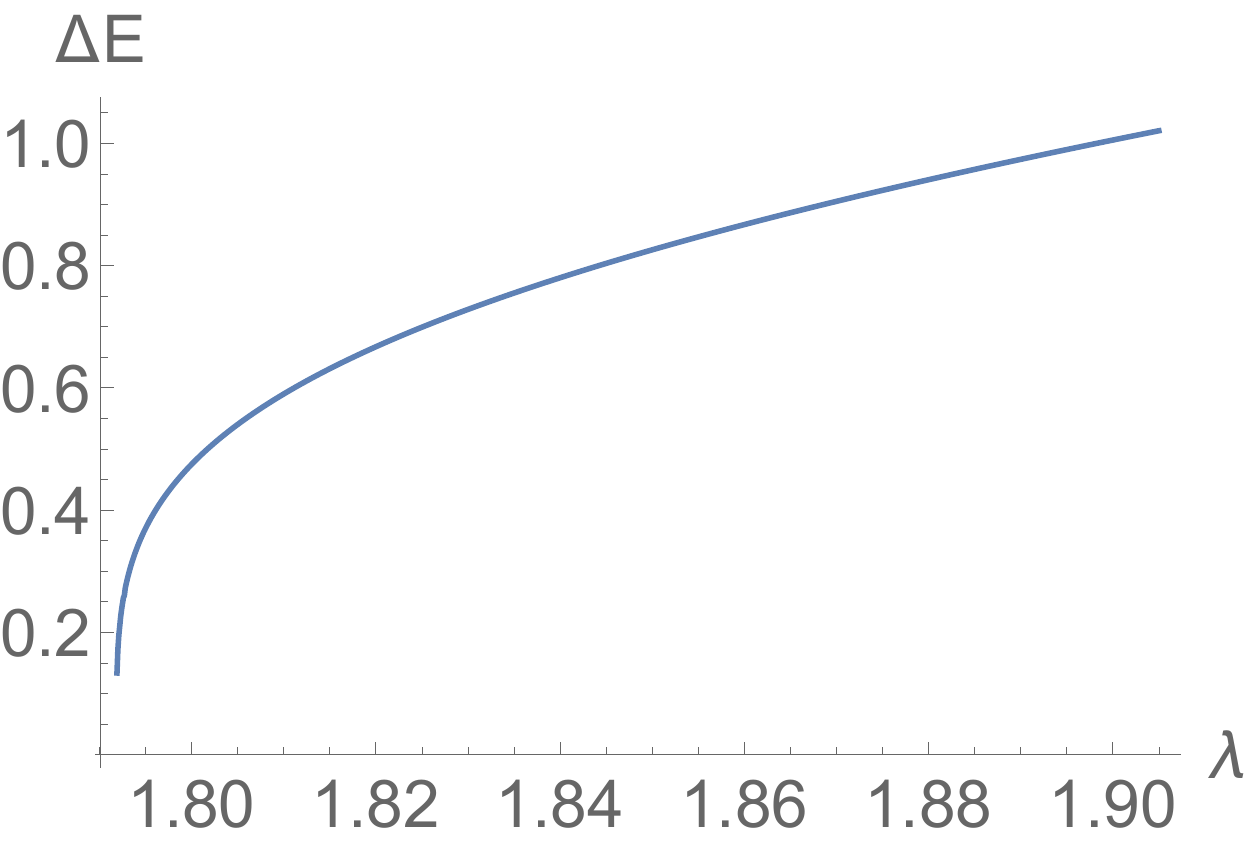}
\caption{Smallest excitation energy of the second-order quantum Hamiltonian  \eqref{quantumHamiltonianSecondOrder}, obtained after Bogoliubov transforming it.}
\label{fig:bogoGap}
\end{subfigure}
\caption{Excitation energy as a function of $\lambda$ for $N \rightarrow \infty$ derived in two different methods. In both cases, we observe a gapless excitation at $\lambda_{lm} \approx 1.792$. Stable excitations only exist for $\lambda \geq  \lambda_{lm}$.
 As is clear from the Hamiltonian \eqref{fullHamiltonian}, the energy unit is $\frac{4\pi^2\hbar^2}{2m L^2}$.}
\end{figure}

Since we already expect  from \fig \ref{fig:HOfX} that a stationary inflection point appears at $\lambda = \lambda_{lm}$, it remains to confirm that this is the case. To this end, we first study the first derivative of the Bogoliubov Hamiltonian. Setting it to zero yields four equations, which we can solve for the four Bogoliubov parameters $x$, $\theta$, $\Delta_2$ and $\Delta_3$. We observe that the latter two parameters behave as in the second minimum, $\Delta_2 = \Delta_3 = 0$. Only the derivatives with respect to $x$ and $\theta$, which are displayed  in equation \eqref{macroscopicHamiltonianFirstOrder} in appendix \ref{app:formulas}, yield non-trivial conditions. 
 As we expect from the previous analysis, solutions, i.e., local extrema, only exist for $\lambda > \lambda_{lm}$, which we determine as $\lambda_{lm} = 1.792$. 
 Next, we compute the matrix $\mathcal{M}$ of second derivatives, which is displayed in equation \eqref{macroscopicHamiltonianSecondOrder}. 
At the local minima, we plug in  the above determined values of the Bogoliubov parameters and then compute its determinant. We display the result in \fig \ref{fig:macroscopicFluctuations} as a function of $\lambda$. We confirm that it vanishes as $\lambda$ approaches $\lambda_{lm}$ from above. Therefore, $\lambda = \lambda_{lm}$ corresponds to a stationary inflection point in the Bogoliubov Hamiltonian and it follows from our $c$-number method that a nearly-gapless mode and consequently an increased degeneracy of states exists in the full spectrum.

In order to confirm this finding, and also to make a more quantitative statement, we explicitly perform the Bogoliubov transformation to obtain the full quantum spectrum in the limit $N \rightarrow \infty$. 
As explained in section \ref{ssec:proof}, the first step is to replace $\hat{a}_1/\hat{a}_1^\dagger \rightarrow \sqrt{N-\hat{a}_2^\dagger \hat{a}_2 - \hat{a}_3^\dagger \hat{a}_3}$
in the full Hamiltonian \eqref{truncatedHamiltonian} to ensure that we only consider fluctuations that respect particle number conservation.
Then we expand to second order around the point defined by the Bogoliubov approximation \eqref{macroscopicReplacements}. We display the result in appendix \ref{app:formulas} in equations \eqref{quantumHamiltonianFirstOrder} and \eqref{quantumHamiltonianSecondOrder}. 
As before, we subsequently look for pairs ($x$, $\theta$) where the linear term \eqref{quantumHamiltonianFirstOrder} vanishes and stable fluctuations exist. We obtain the same values as above.
At those points, we calculate numerically the Bogoliubov transformation for the corresponding quadratic Hamiltonian using the method described in \cite{bogoliubovCalculation1, bogoliubovCalculation2}. The so obtained diagonal matrix contains
the excitation energies associated to the Bogoliubov modes. The smallest energy gap as a function of $\lambda$ is shown in \fig \ref{fig:bogoGap}.
As expected, the result is in accordance with the previous one displayed in \fig \ref{fig:macroscopicFluctuations}. 
Thus, this analysis confirms that a gapless mode exists at $\lambda_{lm}$ in the limit $N\rightarrow \infty$. For finite $N$, we therefore expect that a nearly-gapless mode, whose energy is suppressed as a power of $1/N$, appears close to $\lambda_{lm}$. 

 Finally, we want to point out that this analysis also shows that the critical state, around which the gapless mode emerges, is stable. This follows from the fact that the gaps of all modes are positive and large in the relevant regime $\lambda \geq \lambda_{lm}$, except for one almost flat direction. 
We are interested in the latter because of its importance for information storage.

\subsection{Slow Mode in Full Spectrum}

\subsubsection*{Outline of Approach}

Our goal is to confirm the existence of a light mode in the spectrum of the quantum system for finite $N$. To this end, we will use the fact that modes with a small energy gap $\Delta E$ evolve on the long timescale  $\hbar/\Delta E$ (see the discussion around \Eq \eqref{disturbanceHamiltonian}). When we consider a state close to a critical point of enhanced memory storage, we therefore expect the appearance of large timescales in its time evolution. So we will prove the existence of a light mode by showing that quantum states with a drastically slower time evolution exist there.\footnote
	{For the periodic system, it was shown explicitly that the time evolution is significantly slower at the critical point \cite{goldstone}.}
	As we shall discuss in section \ref{ssec:externalCoupling}, we expect such states to have an experimental signature in the form of absorption lines of low frequency.\footnote{
	We remark that it does not suffice to look for eigenstates in the spectrum whose energies only differ by a small value $\Delta E$. The reason is that even when eigenstates have a similar energy, the transition between them can nevertheless be a suppressed higher-order process, \ie it can be very hard to transit from one to the other. In this case, no light mode exists and a soft external stimulus cannot induce the transition.
	}

The timescale of evolution also determines how long a state can store information. We can imagine that we experimentally prepare a state in such a way that we can choose its components in a certain basis. Then it is possible to encode information in these components. If we measure the state before it has evolved significantly, we can directly read out the components and therefore the stored information. In contrast, if the state has already evolved, it is practically impossible to retrieve the information since this would require precise knowledge of the dynamics of the systems, in particular of its energy levels. In a narrow sense, the timescale of evolution therefore determines a decoherence time. It is the time after which the subset of nearly-gapless modes has been decohered by the rest of the system.

\label{ssec:slowMode}
 Practically, we need to come up with a procedure to single out a quantum state, for which we then analyze its time evolution.
	  For $\lambda = \lambda_{lm}$, this quantum state should correspond to the stationary inflection point of the Bogoliubov Hamiltonian. In order to make a comparative statement, we moreover need to determine analogous quantum states at different values of $\lambda$. We will achieve this by constructing a method to associate a quantum state $\ket{\Phi_{\text{inf}}}$ to the inflection point of the Bogoliubov Hamiltonian. This inflection point exists for all $\lambda \gtrsim 1$ but is stationary only for $\lambda = \lambda_{lm}$.

	 Our approach to determine $\ket{\Phi_{\text{inf}}}$ is to define a subspace of states close to the inflection point and then to select among those the state of minimal energy. On the one hand, this subspace should not be too big in order to be sensitive to properties of the stationary inflection point at $\lambda = \lambda_{lm}$. On the other hand, the subspace cannot be too small since otherwise the energy of the state that we obtain by minimization is too high. Of course, there is no unique way to determine this subspace and therefore no unique quantum state $\ket{\Phi_{\text{inf}}}$, but it suffices for us to come up with a method to find some quantum state with drastically slower time evolution. In particular, we expect that there are many different such quantum states corresponding to different occupation numbers of the light mode.
	 
	 \subsubsection*{Concrete Procedure}
	
 Concretely, we will construct the subspace using two conditions, which we derive from properties of the inflection point $x_{\text{inf}}(\lambda)$ of the Bogoliubov Hamiltonian. First, we only consider quantum states $\ket{\Phi_{\text{inf}}}$ for which the expectation value of the relative occupation of the 2-mode, $n_2(t) :=\bra{\Phi_{\text{inf}}} \hat{a}_2^\dagger(t) \hat{a}_2(t)\ket{\Phi_{\text{inf}}}/N$, is equal to $x_{\text{inf}}(\lambda)$. Secondly, we restrict the basis used to form the quantum state $\ket{\Phi_{\text{inf}}}$, where -- as in all numerical computations -- we use number eigenstates of total occupation number $N$ as basis. Namely, we determine from the Bogoliubov Hamiltonian all relative occupation numbers at the inflection point. Then we choose upper bounds $\delta n_i$ on the spread of the different modes, i.e, we only consider basis elements for which the relative occupation numbers deviate by at most $\delta n_i$ from the values determined from the Bogoliubov Hamiltonian. With the guideline that modes with bigger relative occupation should have a bigger spread, we empirically determine $\delta n_1 = 0.4$, $\delta n_2 = 0.375$ and $\delta n_3 = 0.225$ to be a good choice.\footnote
 {We compared the results obtained in this truncation with the ones derived using the full basis. For $N\leq 50$, we observed that their qualitative behavior, which we shall discuss in a moment, is identical whereas this no longer seems to be the case for higher $N$. However, the only important point for us is to come up with some recipe to find the slowly evolving states.}

\begin{figure}
	\begin{center}
		\begin{subfigure}{0.55\textwidth}
			\centering
			\includegraphics[scale=0.6]{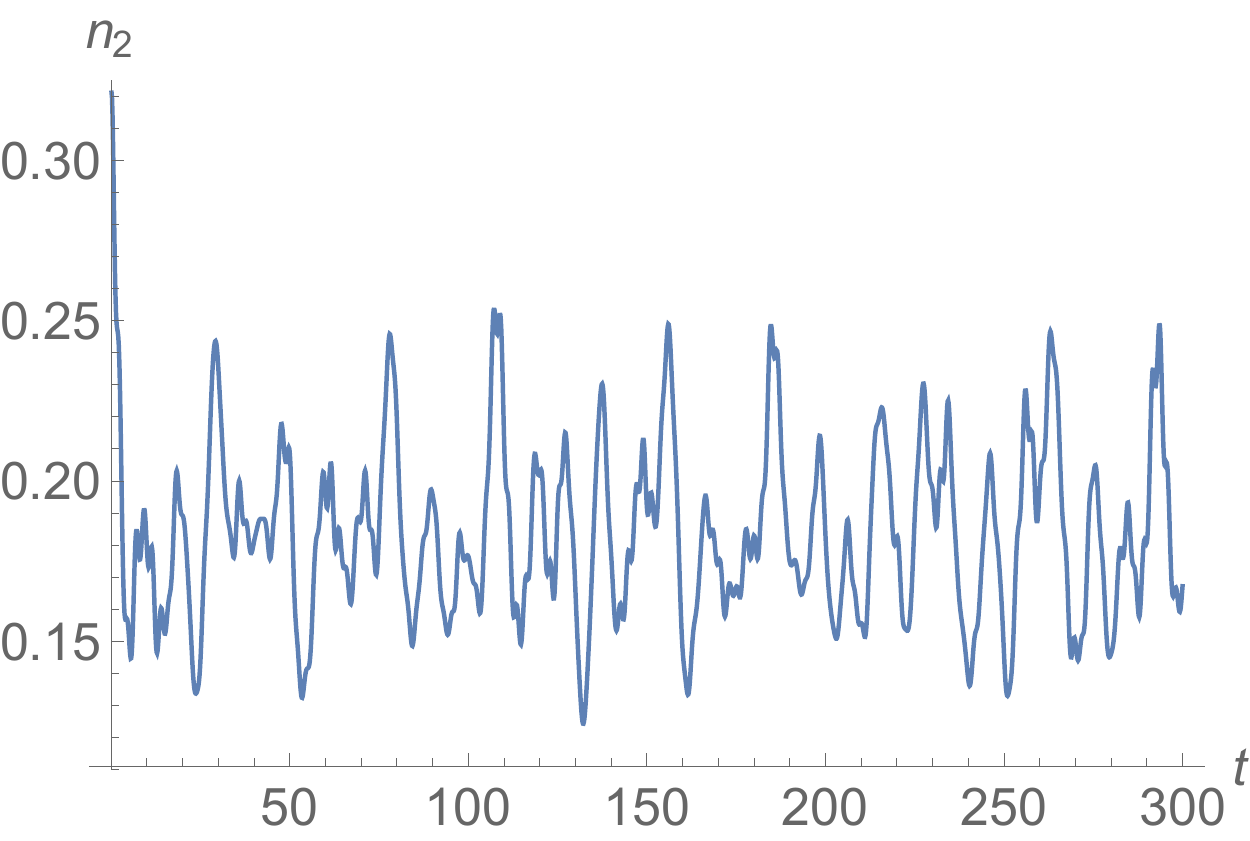}
			\caption{$\lambda = 1.9$}
			\label{fig:slowModesTimea}
		\end{subfigure}
		\begin{subfigure}{0.55\textwidth}
			\centering
			\includegraphics[scale=0.6]{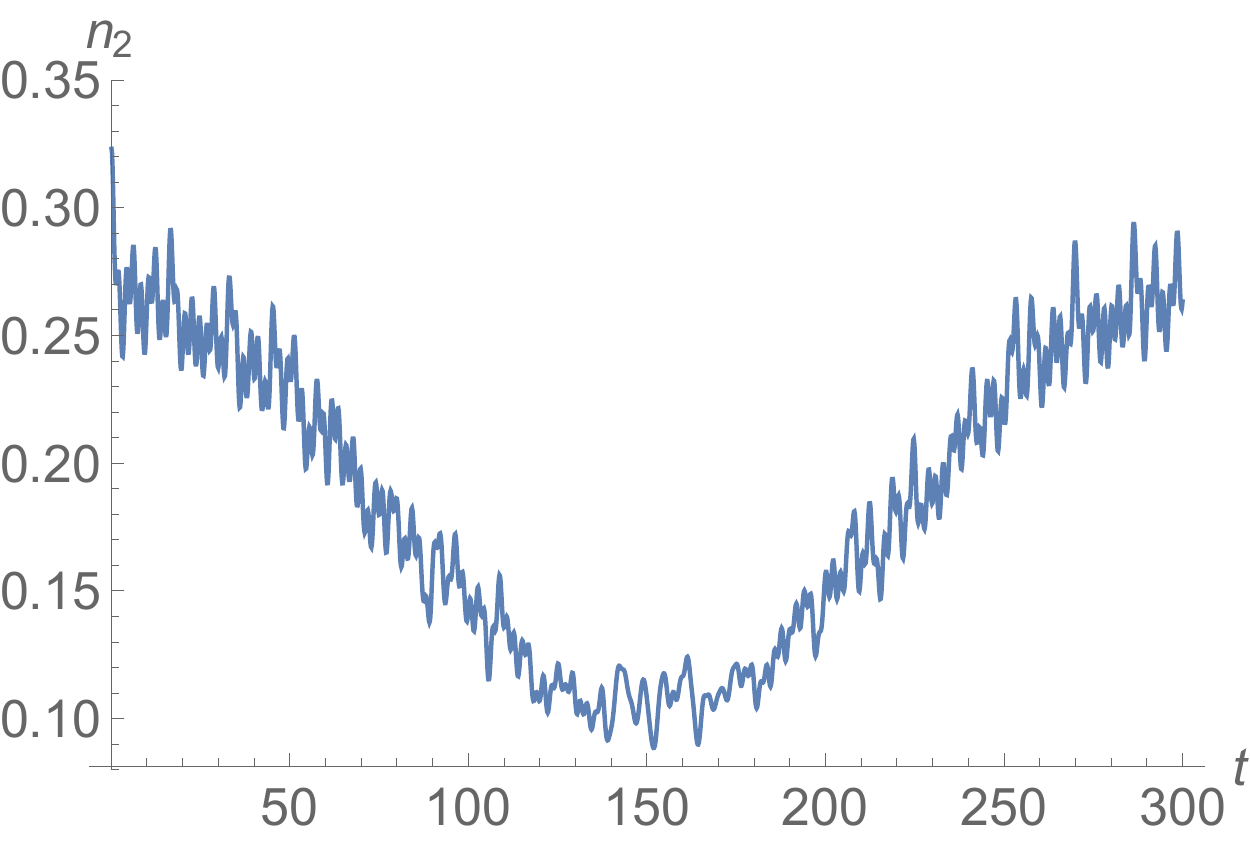}
			\caption{$\lambda = 2.083$}
			\label{fig:slowModesTimeb}
		\end{subfigure}
		\begin{subfigure}{0.55\textwidth}
			\centering
			\includegraphics[scale=0.6]{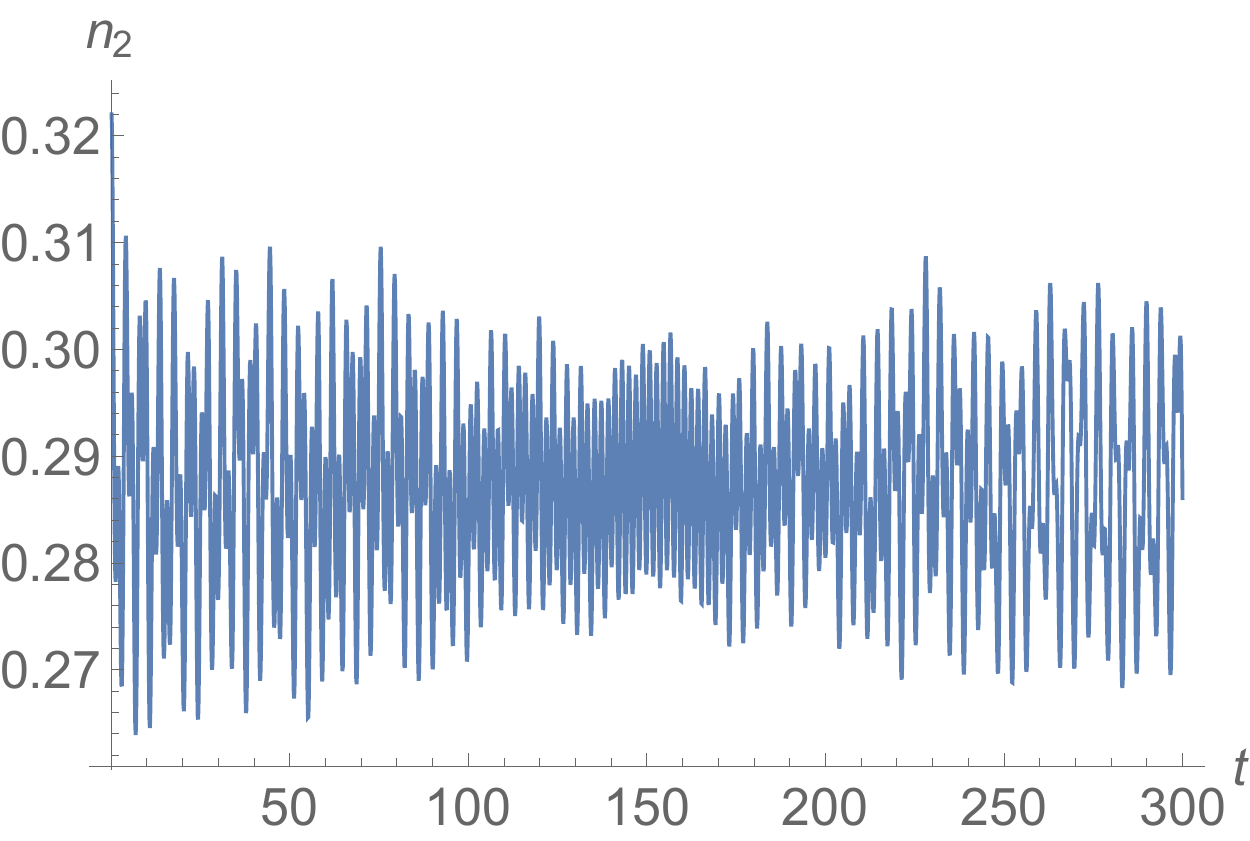}
			\caption{$\lambda = 2.2$}
			\label{fig:slowModesTimec}
		\end{subfigure}
		\caption{Time evolution of the quantum state $\ket{\Phi_{\text{inf}}}$, which corresponds to the inflection point of the Bogoliubov Hamiltonian. The value of $n_2(t)$ is plotted for $N=60$. We observe that lower frequencies dominate around $\lambda \approx 2.083$.}
		\label{fig:slowModesTime}
	\end{center}
\end{figure}

\begin{figure}
	\begin{center}		
		\includegraphics[scale=0.7]{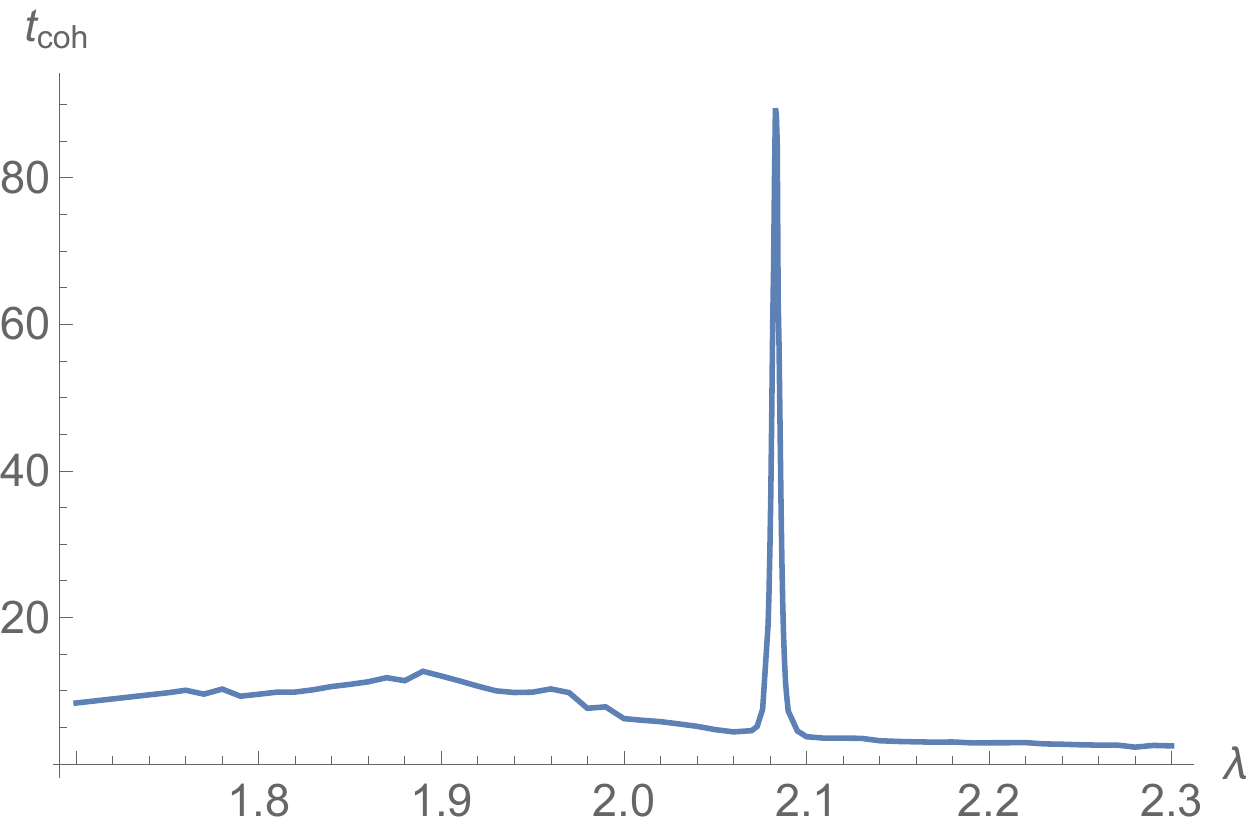}
		\caption{Estimate of the decoherence time $t_{\text{coh}}$ associated to $\ket{\Phi_{\text{inf}}}$ as a function of $\lambda$ for $N=60$. We observe  that it increases distinctly around $\lambda\approx 2.083$.}
		\label{fig:meanFrequency}
	\end{center}
\end{figure}

Once we have determined the state $\ket{\Phi_{\text{inf}}}$, we study its time evolution. In doing so, we use the full quantum Hamiltonian and therefore also the full basis. We show the result for $N=60$ for exemplary values of $\lambda$ in \fig \ref{fig:slowModesTime}, where we plotted $n_2(t)$. Clearly, drastically lower frequencies dominate around $\lambda = \lambda_{lm}$. 
In order to make a quantitative estimate about the coherence time as a function of the collective coupling $\lambda$, we extract a typical frequency from the time evolution of $\ket{\Phi_{\text{inf}}}$.
To this end, we use a discrete Fourier transformation with respect to the $n_{\text{max}}$ frequencies $f_1, 2f_1, \ldots, n_{\text{max}} f_1$
to obtain the Fourier coefficients $c_1, c_2, \ldots, c_{n_{\text{max}}}$. With their help, we can define the mean frequency as
\begin{equation}
	\bar{f} :=  f_1 \frac{\sum_{i=1}^{n_{\text{max}}} i |c_i|^2 }{\sum_{i=1}^{n_{\text{max}}} |c_i|^2} \,.
	\label{meanFrequency}
\end{equation}
As explained, the timescale of evolution can be interpreted as decoherence time, namely as the timescale after which the subset of nearly-gapless modes has been decohered by the other modes. In this sense, we get:
\begin{equation}
t_{\text{coh}} = \frac{1}{\bar{f}} \,.
\end{equation}
For $f_1 = 1/3000$ and $n_{\text{max}}=12000$, we show $t_{\text{coh}}$ as a function of $\lambda$ in \fig \ref{fig:meanFrequency}.\footnote
{We checked that different choices of $f_1$ and $n_{\text{max}}$ lead to the same result. Therefore, cutting off low and high frequencies, which is required in a numerical treatment, has no influence on our findings.}
We observe that it increases distinctly around $\lambda \approx 2.083$.

\begin{figure}
 	\begin{center}
 		\includegraphics[scale=0.3]{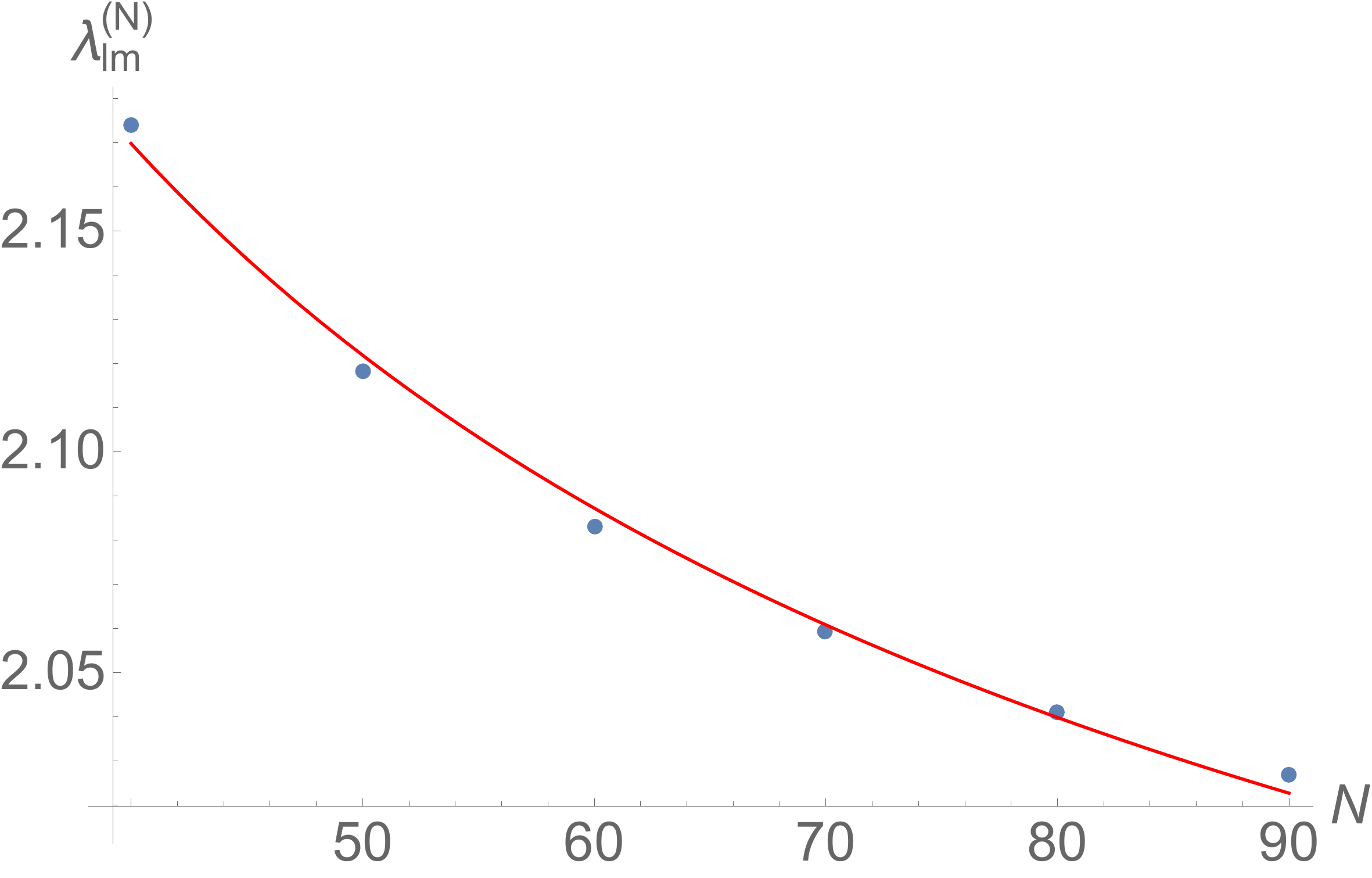}
 		\caption{ Critical value $\lambda^{(N)}_{lm}$ as a function of particle number $N$. The positions obtained from numerical simulations are plotted in blue. The fitted function \eqref{lambdaFit} is shown in red.}
 		\label{fig:lambdaVsN}	
 	\end{center}
 \end{figure}

The fact that for $N=60$ a state with drastically slower time evolution appears at $\lambda^{(60)}_{lm} \equiv 2.083$ is consistent with $\lambda_{lm} = 1.792$ since we expect that as in the periodic system, the critical value $\lambda^{(N)}_{lm}$ of the collective coupling at finite $N$ receives $1/N$-corrections:\footnote
{In contrast, note that we do not expect $t_{\text{coh}}$ to diverge for infinite $N$ because $\ket{\Phi_{\text{inf}}}$ generically contains an admixture of non-gapless modes.}
\begin{equation}
\lambda^{(N)}_{lm} = \lambda_{lm} + a \cdot N^{-b},
\label{lambdaFit}
\end{equation}
where $a>0$ and $b>0$ are two undetermined parameters. 
 To confirm that this is the case, we repeated the above analysis of the decoherence time as a function of $\lambda$ for $N$ between $40$ and $90$.  
This determines critical values $\lambda^{(N)}_{lm}$ as the values of $\lambda$ for which time evolution is the slowest at a given $N$.
Subsequently, we fit the function \eqref{lambdaFit} to the result and thereby determine $a=3.56$ and $b =0.61$. We note that $b$ is close to $2/3$, which was the result in the periodic system \cite{kanamoto}. As is evident from \fig \ref{fig:lambdaVsN}, the numerically determined values $\lambda^{(N)}_{lm}$ are well described by the fitted function \eqref{lambdaFit}.
This is a clear indication that the slowed time evolution we found is due to the nearly-gapless Bogoliubov mode that we predict from the analytic treatment. So in summary, we observe the appearance of a nearly-gapless mode around $\lambda = \lambda_{lm}$ also for finite $N$. 

\begin{figure}
	\begin{center}
		\includegraphics[scale=0.3]{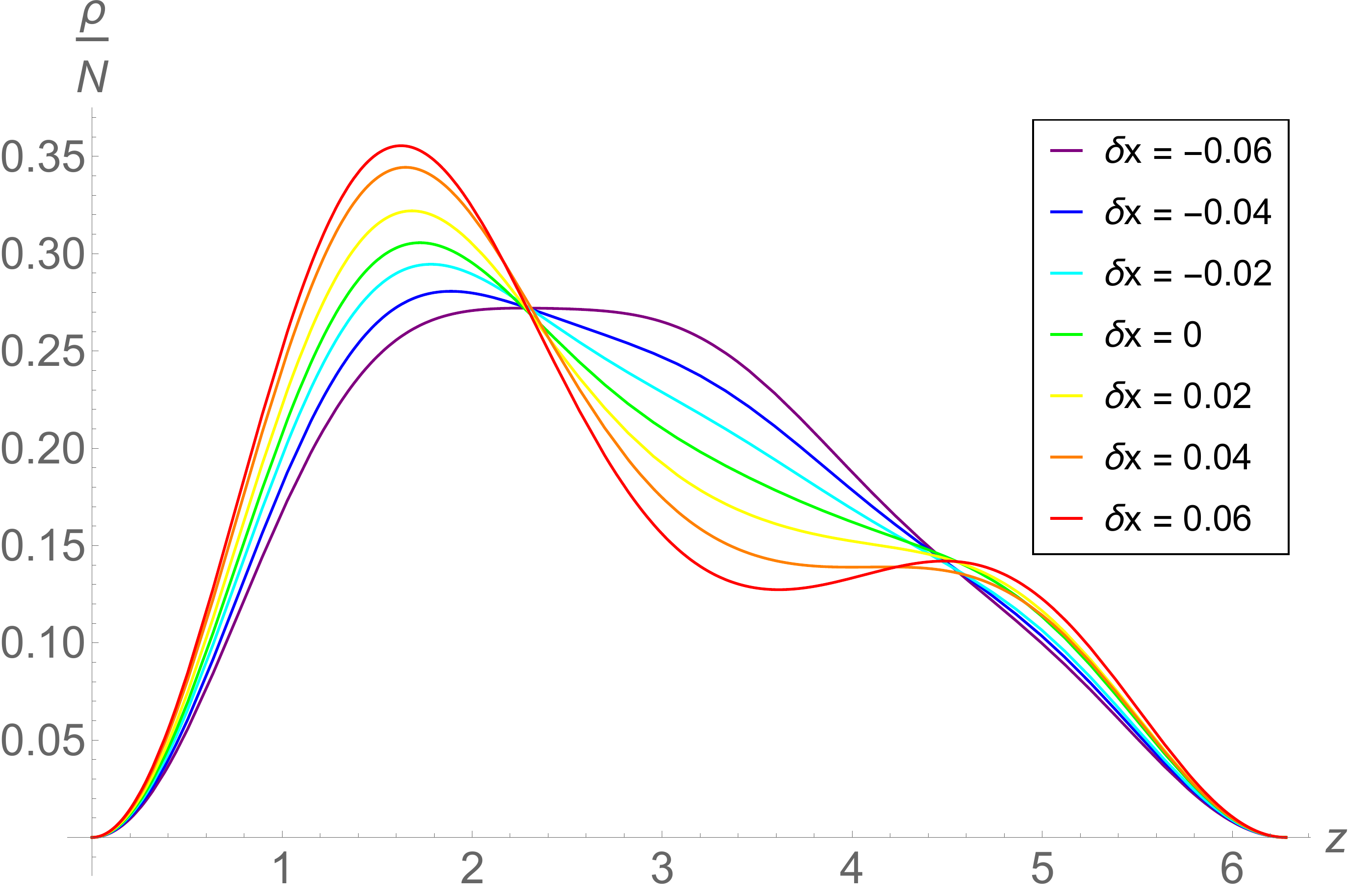}
		\caption{Variations of the critical state at $\lambda=2.083$ for $N=60$ in position space. The relative particle density $\rho/N$ is plotted. The green line corresponds to the critical state $\ket{\Phi_{\text{inf}}}$ itself and the adjacent lines are variations of it, which we obtained by slightly changing the value of $x$ used in the minimization procedure that determines the quantum state: $x_i=x_{\text{inf}}(\lambda) + \delta x_i$. The values of $\delta x_i$ are indicated in the plot.}
		\label{fig:positionSpace}	
	\end{center}
\end{figure}

 As a final remark, we discuss the critical state $\ket{\Phi_{\text{inf}}}$ for $N=60$ and $\lambda =2.083$ in position space. Its particle density is given by
\begin{equation}
	\rho(z)\equiv	\bra{\Phi_{\text{inf}}} |\hat{\psi}|^2 \ket{\Phi_{\text{inf}}} = \frac{1}{\pi}\sum_{k,l=1}^3 \bra{\Phi_{\text{inf}}} \hat{a}^\dagger_k \hat{a}_l\ket{\Phi_{\text{inf}}} \sin\left(\frac{kz}{2}\right) \sin\left(\frac{lz}{2}\right)  \,.
\end{equation}
We display it in \fig \ref{fig:positionSpace}, where we also illustrate what the gapless mode looks like in position space. To this end, we fix the critical value of the collective coupling, $\lambda =2.083$, but slightly vary the value of $x$ used in the minimization procedure that determines the quantum state: $x_i=x_{\text{inf}}(\lambda) + \delta x_i$. This determines a family of quantum states $\ket{\Phi_{\text{inf,\ i}}}$, where $\ket{\Phi_{\text{inf,\ i}}}$ is a state of minimal energy subject to the constraint that its relative occupation of the 2-mode is $x_i$. Their particles densities are also shown in \fig \ref{fig:positionSpace}. 

\subsection{Comparison with Goldstone Phenomenon} 
 
 It may be useful to compare our effect with the well-known phenomenon
 of appearance of gapless excitations in the form of Goldstone bosons.  
The latter modes emerge as a result of a phase transition with the spontaneous breaking of a global symmetry.  The crucial difference 
is that Goldstone modes consistently exist in a domain past 
the critical phase. This is not the case in the present model. 
Our gapless modes only exist at the critical point and they appear due to 
cancellation between the positive kinetic energy and a negative 
 collective interaction energy with a certain highly-occupied  master mode.  It is therefore hard to interpret the appearance of our gapless modes in terms of a Goldstone phenomenon of spontaneous breaking of any global symmetry.  This difference is what in particular  makes the phenomenon of assisted gaplessness interesting since there is no {\it a priori}  symmetry reason for the emergence of any gapless modes.
 This said, however, once assisted gaplessness takes place, the number of such modes can be highly enhanced by  {\it unbroken}  symmetries    
of the critical state, such as spherical symmetry \cite{areaLaw}.

\subsection{Encoding Information via Coupling to an External Field}
\label{ssec:externalCoupling}

We conclude by pointing out how the nearly-gapless mode emerging at the critical value $\lambda_{lm}$ of the collective coupling can be probed by an external field. 
This is of particular interest with regard to the experimental realization of an attractive Bose gas at the critical point. For simplicity, we will only consider the nearly-gapless mode of the Bose gas, which we would obtain after a Bogoliubov transformation of the Hamiltonian \eqref{truncatedHamiltonian}. We will call it $\hat{b}$ and its gap $\Delta E$. In the vicinity of the critical point, it is possible to neglect the effect of all other modes of the Bose gas if the energy of the external mode is sufficiently small. Following \cite{mischa3}, the essential features  of the coupling to an external field can be captured by the simple Hamiltonian\footnote
{In the expansion of the Hamiltonian around the critical point, the coupling in \eqref{coupledHamiltonian} is expected to be the leading order term coming from the interaction term $\hat{a}^\dagger \hat{a}(\hat{c} + \hat{c}^\dagger)$ of the original $\hat{a}$-modes, which conserves their number and momentum.}
\begin{equation}
	\hat{H}_{\text{eff}} = \Delta E \hat{b}^\dagger \hat{b} + E_\gamma \hat{c}^\dagger \hat{c} + \frac{g}{2} \left(\hat{b}\hat{c}^\dagger + \hat{b}^\dagger \hat{c}\right) \,,
	\label{coupledHamiltonian}
\end{equation}
where we added an external mode $\hat{c}$ of energy $E_\gamma$, which could correspond to a mode of the photon field in an experimental setup. The parameter $g$ describes the coupling of the external mode to the Bose gas. Note that the model \eqref{coupledHamiltonian} is a special case of the Hamiltonian \eqref{disturbanceHamiltonian}, which describes the general coupling of a nearly-gapless mode to an external one.

We can start with no excitations in the Bogoliubov mode and a coherent state of the external field:
\begin{equation}
	\ket{\Phi(t=0)} = \ket{0}_b \otimes \ket{\gamma}_c \,,
\end{equation} 
where $\gamma$ parameterizes the occupation of the external coherent state. Straightforward calculation gives \cite{mischa3}: 
\begin{equation} \label{evolutionExternal}
	\bra{\Phi(t)} \hat{c}^\dagger \hat{c} \ket{\Phi(t)} = \gamma^2\left( 1- \frac{g^2}{\delta_g^2}\sin^2\left(\frac{\delta_g t}{2}\right) \right) \,,
\end{equation}
where we defined
\begin{equation}
	\delta_g = \sqrt{(E_\gamma - \Delta E)^2 + g^2} \,.
\end{equation}
Thus, the coupling to the critical system leads to a fluctuating occupation of the external field. One has to choose $E_\gamma \approx \Delta E$ to allow for an appreciable amplitude of the oscillation. As discussed in the context of \Eq \eqref{disturbanceHamiltonian}, this implies that the coupling has to be small, $g\lessapprox \Delta E$, in order not to disturb the gaplessness of the mode $\hat{b}$.
	
In this situation, the occupation number of the 
 Bogoliubov mode evolves in time as,
\begin{equation} \label{evolutionBogoliubov}
	\bra{\Phi(t)} \hat{b}^\dagger \hat{b} \ket{\Phi(t)} = \gamma^2\, \frac{g^2}{\delta_g^2}\sin^2\left(\frac{\delta_g t}{2}\right) \,. 
\end{equation}
This means that we can use soft excitations of an external field to bring the Bogoliubov mode to a desired state. Correspondingly, information can be read out from soft  $\hat{c}$-quanta that are emitted due to the de-excitement of the Bogoliubov mode.
We remark that equations \eqref{evolutionExternal} and \eqref{evolutionBogoliubov} also show why soft external radiation is sensitive to the critical point. If we are not at the critical point $\lambda = \lambda_{lm}$ of light modes, the lightest mode has a much bigger gap, i.e., $\Delta E \gg E_\gamma$. In that case, the amplitude of fluctuations gets suppressed as $g/\Delta E$. This means that soft radiation stops interacting with the system. Thus, a critical point exists whenever soft radiation, whose energy $E_\gamma$ is much smaller than the typical energy of the system, i.e., $E_\gamma \ll\hbar^2/(2m L^2)$, interacts significantly with the atoms.

The energy efficiency of the information storage within the gapless mode goes hand in hand with the difficulty of the read-out of information. 
Since the gap is small, the different information patterns are barely discriminable and the read-out time is correspondingly very long. 
 From the theoretical perspective of understanding black holes this 
 is good news as such a delay would naturally explain why the quantum information 
 stored in black hole modes cannot be resolved for a very long time.    
 The answer to the above black hole puzzle would be that information
is unreadable because it is stored in nearly gapless modes 
(see e.g., \cite{mischa3}). 
 
 Regarding a possible practical use of assisted gaplessness for the storage 
 and read-out of quantum information, the same constraints apply. 
 Namely, if we would like to design a device that could read out the 
 information on a timescale shorter than the inverse gap, such   
 a reader must necessarily disturb the gap. In such a case, the reading device can be included in an effective Hamiltonian in form of a 
 time-dependent interaction term that is switched on externally
 when needed.  We shall not elaborate further on this issue in the present paper.
   
  Finally, we want to comment on the stability of the critical state, around which the gapless mode emerges.  Of course, as we have explained,  this state is not a ground state 
of the system. Nevertheless, it exhibits no Lyapunov exponent in any 
direction of the energy landscape.  This is most clearly demonstrated  by the analytic studies performed in the limit of large $N_0$. In particular, plot \ref{fig:bogoGap}  
of the gap of the lowest-lying Bogoliubov mode as a function of 
$\lambda$ shows that in the relevant regime $\lambda \geq \lambda_{lm}$, 
all the gaps are positive and only one becomes zero at criticality. 
Moreover, we see no signs of any decay in the full numerical analysis of the system, which again confirms the absence of any instability.  
Correspondingly, in the closed system with Hamiltonian  
\eqref{truncatedHamiltonian} the critical state is stable. 

 Of course, the system can be destabilized by coupling it to external
 modes provided the coupling is strong enough. But as we have discussed, any interaction to an external field has to be weak anyhow in order not to disturb the gaplessness of the Bogoliubov mode. We expect that the weakness of the external influence also ensures a sufficient stability of the critical state, although the matter has to be studied on a case by case basis
for potential experimental setups. 
If the above requirements can be met, it would be highly interesting to scatter an electromagnetic wave at the system of cold atoms that are in the critical state and to see if it is experimentally possible to  detect  the emission of soft  radiation resulting from the de-excitement of the Bogoliubov mode. In this way, the quantum information stored in 
the memory in the form of an excited state of the Bogoliubov mode could be decoded by analyzing the emitted soft radiation. 
 
\section{Implications for Neural Networks}
\label{sec:neural}
\subsection{Mapping of Bosonic System on Neural Network}
We shall now show that the enhanced memory capacity phenomenon discussed above has a direct application to quantum neural networks
along the lines of \cite{neural1,neural2}. 
First, we will introduce how neural networks can be described by an effective Hamiltonian and then we will specialize to our prototype system \eqref{truncatedHamiltonian}, both on the quantum level and in the classical limit.

The key ingredients of any neural network are on the one hand the neurons and on the other hand the synaptic connections among them. Following \cite{neural1,neural2}, it is possible to describe them by an effective Hamiltonian in which the neural excitations are the degrees of freedom. The threshold excitations correspond to kinetic energies and the synaptic connections translate as interaction terms. In this description, the time evolution of the excitations is generated by the effective Hamiltonian. In particular, this framework enables us to study the energetics of information storage. This is the aspect we shall focus on. So we will not study any specific algorithm, but our goal is to investigate the energy cost of recording and reading out information.

The synaptic connections can be either excitatory or inhibitory, i.e., an excitation of a given neuron $k$ can either decrease or increase the probability of the excitation of another neuron $j$. 
In the effective Hamiltonian description of the network,
the excitatory and inhibitory nature of the connections can be given an energetic meaning. This meaning is defined by the sign of the interaction 
 energy of two or more simultaneously excited neurons. 
 The negative and the positive signs of the interaction
 energy  respectively translate as excitatory and inhibitory   
connections in an {\it energetic sense}.  
 Below, we shall denote the parameter that sets the characteristic strength of these interactions 
 by the same constant $\alpha$ as we have used for the system of bosons. 

 As noticed in  \cite{neural1,neural2}, a 
 neural network defined in this way in the case of 
 negative  (i.e., attractive and therefore "gravity-like") synaptic connections 
 exhibits the phenomenon of assisted gaplessness with sharply enhanced   
 memory storage capacity.
For describing the idea it suffices to consider the simple Hamiltonian \eqref{FH1} and to think of it as representing a quantum neural network.  Thus, we consider a network for which the synaptic connection energy of a set of inter-connected neurons is negative. 
 This means that the excitation of a given neuron lowers the threshold for 
the excitation of all the other neurons from this set, i.e., 
for the ones that are connected to the former neuron
by negative energy couplings. 
If we normalize the characteristic 
step of the excitation to unity, then exciting a neuron to 
a level $N$, in general,  lowers the threshold for the others by the amount $\sim \alpha N$. 

  Due to this effect, there  generically exist states in which the high excitation of a subset of neurons assists other neurons in becoming gapless. Their gap is lowered to zero up to an accuracy of order $\alpha$.  Correspondingly, we encounter the interesting situation that the effect becomes stronger for weaker coupling. Namely we can shrink the gap by decreasing the coupling strength $\alpha$ and simultaneously increasing the excitation level $N$ in such a way that their product $\alpha N$ is constant.
   In such states, the memory storage in the emerging gapless neurons becomes energetically very cheap and one can store a large number of patterns within 
a very narrow energy gap.  By taking the double-scaling limit as defined in \eqref{doubleScaling}, the gap can be made arbitrarily narrow and the number of stored patterns arbitrarily large. 

 Furthermore, it was shown in \cite{neural2}  that such a neural network is  isomorphic to the physical system of  a bosonic quantum field.   In this correspondence, 
 the neural degrees of freedom are identified with 
 the momentum modes of the field, whereas the synaptic connections correspond to the couplings among the different momentum modes. This mapping allows to give a unified description of the 
 phenomena of enhanced memory storage capacity
 in  neural networks and in systems with cold bosons.   This opens up an exciting experimental prospect of simulating such neural networks in a system of cold atoms. 

\begin{figure}
	\begin{center}
		\includegraphics[scale=0.6]{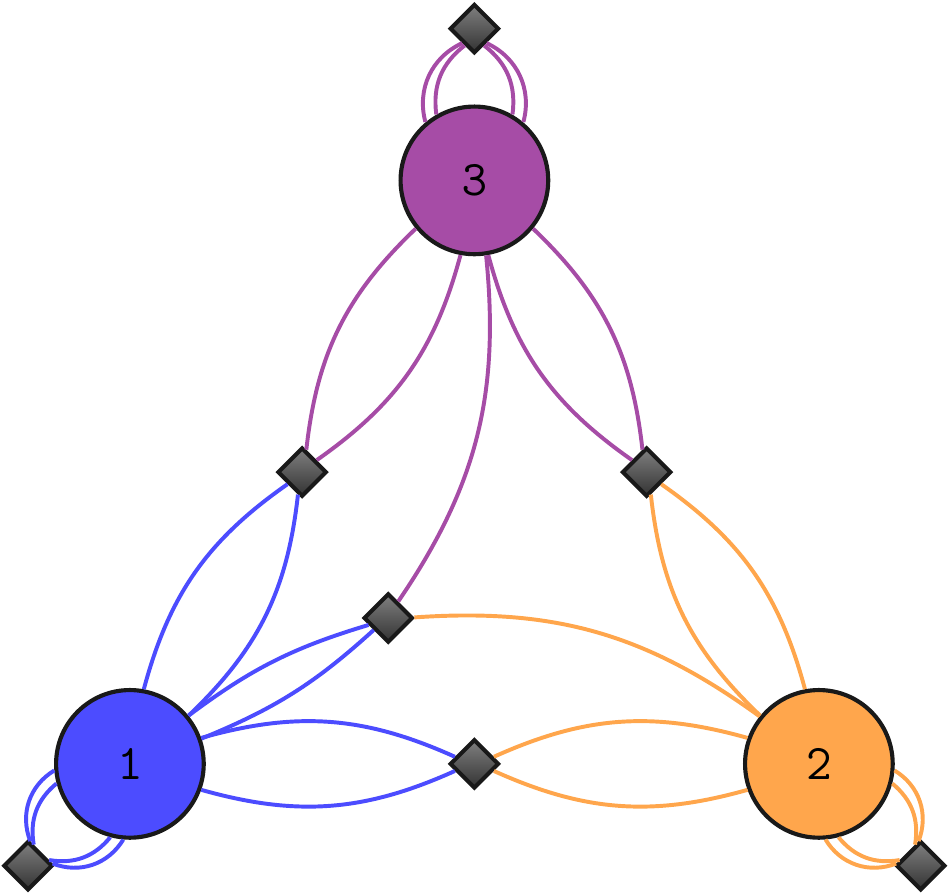}
		\caption{ Representation of the Hamiltonian \eqref{truncatedHamiltonian} as neural network. The three neurons are displayed as circles and diamonds represent interaction terms. The number of lines to a diamond indicates how many mode operators of the corresponding neuron participate in the interaction.}
		\label{fig:neural}	
	\end{center}
\end{figure}

  Using this dictionary, we shall represent the system studied in the present paper as a neural network.  
Indeed, the truncated Hamiltonian (\ref{truncatedHamiltonian}) is fully isomorphic 
to a quantum neural network with three neurons in which the excitations of neurons are  
described by the momentum modes $\hat{a}_k^\dagger, \hat{a}_k$, and the synaptic connections between the neurons are described by the interaction 
terms in (\ref{truncatedHamiltonian}).\footnote
{We remark that this system only represents a part of a neural network that can be used for actual memory storage. In a realistic situation, one needs a first device to input information, a second one to store it and a third one to retrieve it. However, our system solely realizes the second part. So we only study situations in which no input or output operations take place.}
\fig \ref{fig:neural} shows the representation of Hamiltonian \eqref{truncatedHamiltonian} as neural network.

 To make a closer contact with the neural 
network language, it is useful to rewrite the 
Hamiltonian as 
\begin{equation} \label{HNET}
\hat{H} =   \sum_{k=1}^3 E_k \hat{a}_k^\dagger \hat{a}_k 
 - \sum_{k,j=1}^3 \hat{a}_k^\dagger\hat{{\mathcal W}}_{kj} \hat{a}_j \,,
 \end{equation} 
 where $E_k = \frac{1}{4} k^2$ are the threshold excitation energies of the neurons and 
   $\hat{{\mathcal W}}_{kj}$ is a Hermitian $3\times 3$ matrix operator of synaptic connections. Its elements are: 
   \begin{subequations}  
\begin{eqnarray}
\hat{{\mathcal W}}_{11} & = & \frac{3\alpha}{8} \hat{a}^{\dagger}_1\hat{a}_1 \,,\\
\hat{{\mathcal W}}_{22} & = & \frac{3\alpha}{8} \hat{a}^{\dagger}_2\hat{a}_2 \,,\\
\hat{{\mathcal W}}_{33} & = & \frac{3\alpha}{8} \hat{a}^{\dagger}_3\hat{a}_3 \,,\\
\hat{{\mathcal W}}_{12} & = & \frac{\alpha}{8} 
\left(4\hat{a}^{\dagger}_2\hat{a}_1  + 2\hat{a}^{\dagger}_1\hat{a}_2 + {4 \over 3} \hat{a}^{\dagger}_2\hat{a}_3 + \hat{a}^{\dagger}_3\hat{a}_2\right ) \,,\\
\hat{{\mathcal W}}_{13} & = & \frac{\alpha}{8} \left(
4\hat{a}^{\dagger}_3\hat{a}_1  + 2\hat{a}^{\dagger}_1\hat{a}_3 + {4 \over 3} \hat{a}^{\dagger}_2\hat{a}_2 - 2 \hat{a}^{\dagger}_1\hat{a}_1\right ) \,,\\
\hat{{\mathcal W}}_{23} & = & \frac{\alpha}{8}  \left (
4\hat{a}^{\dagger}_3\hat{a}_2  + 2\hat{a}^{\dagger}_2\hat{a}_3 + {4 \over 3} \hat{a}^{\dagger}_1\hat{a}_2 + \hat{a}^{\dagger}_2\hat{a}_1\right ) \,.
\end{eqnarray}
\end{subequations}
Now we can directly apply all our results to the 
above neural network. We shall all the time assume the regime of very weak synaptic connections, i.e.,
we assume  $\alpha \ll 1$.  In this situation, we 
study the memory storage capacity of the network 
for various values of the total excitation level $N$.   

\subsection{Enhanced Memory Storage}
First, we examine the memory storage capacity
of the neural network around a state in which
the total excitation level is well below the critical level, 
$N \ll {1 \over \alpha}$.
 In such a regime, the negative 
energy of synaptic connections is negligible
and does not contribute to lowering the energy gap.   
Note that  the ability of the synaptic connection energy to lower the gap of neurons  is parameterized by the 
strength of the collective coupling $\lambda= \alpha N$, which 
is very weak in the considered regime, $N \ll {1 \over \alpha}$.
 
Correspondingly, in such a regime the energy difference between different number eigenstates 
$\ket{N-n_2 -n_3, n_2,n_3}$ and 
$\ket{N-n_2' -n_3', n_2',n_3'}$ mostly comes from 
the first threshold energy term in the network Hamiltonian 
(\ref{HNET}) and is very large
\begin{equation}\label{DIFFbig} 
  \Delta E_{\lambda\ll 1}  = {1 \over 4} \left( 
  3 (n_2' - n_2) + 8(n_3' -n_3) \right) + 
  {\mathcal O}(\lambda) \, .
  \end{equation}  
Hence, the patterns stored in such states 
occupy a very large energy gap. 
 For example, in order to redial a pattern stored in 
 the state $\ket{N,0,0}$ into the state
 $\ket{N-1,1,0}$, we need to overcome an energy gap $\Delta E \simeq {3 \over 4}$, i.e, an external stimulus that is needed for the redial of information 
 $\ket{N,0,0} \rightarrow \ket{N-1,1,0}$
 has to have an energy of order 
$ \sim {3 \over 4}$. 

Now we increase the total excitation level 
 $N$. With this increase, the contribution of 
 the negative synaptic connection energy gradually lowers the gap between some neighboring states. 
  The gap reaches the smallest value when the total excitation level $N$ reaches the  critical point.
 As discussed above, this happens when 
 $ \alpha N =\lambda_{lm}$.  
At this point, a 
 state becomes available around which a
 gapless excitation emerges. This means that for a certain
 relative distribution of excitation levels, the 
 gap between the set of patterns collapses to 
\begin{equation} \label{DIFFsmall}
 \Delta E_{\lambda = \lambda_{lm}} \sim {1\over N^\beta} \,,
 \end{equation}
 where $\beta$ is a positive constant.
  By taking the double scaling limit \eqref{doubleScaling}, we can make the gap arbitrarily narrow.  
  In this situation, the storage of patterns in such states becomes energetically cheap. 
 
 So far, we have described a quantum neural network. We can move to a classical neural network by
 using the coherent state basis and replacing 
 the Hamiltonian operator by its expectation value over the coherent states.   
 In this way, we obtain a classical neural network 
 described by the $c$-number energy function
 $H_{\text{bog}}(\vec{a},\vec{a}^*)$.  
 
  When discussing the storage of patterns in a neural 
 network, we must introduce the notion of the pattern 
 vector. In general, this vector is different from the quantum state 
 vector of the system and may contain less information.  
This is determined by those characteristics of the state to which an external reading device is sensitive.  
  Indeed, when the underlying quantum state characterizing
  the system is given by a coherent state, which 
  is labeled by three complex numbers 
  $\ket{a_1,a_2, a_3}$, the storage of a pattern and therefore the pattern vector
  is determined by the combinations of these numbers to which the
 reader is sensitive.  
  If the reader is sensitive to the full 
 quantum information, i.e., to  
 the phases as well as to the absolute values of $a_j$, then  
  the pattern vector can be identical to the quantum-state vector.  If instead the reader is only sensitive 
  to the absolute values, the pattern vector 
  can be   correspondingly chosen in the form 
  $(|a_1|, |a_2|, |a_3|)$.   Since we do not specify any external reading device, we shall keep the definition of the pattern vector flexible. 
 
We can explicitly write down the pattern vector for our concrete system \eqref{HNET}. First, we discuss the case in which the reader is sensitive to the full quantum information. Using the notations \eqref{macroscopicReplacements},
the pattern vector can then be parameterized as 
 \begin{equation}  \label{patternVector}
 \begin{pmatrix}
 a_1   \\
 a_2  \\
 a_3
 \end{pmatrix}
 = \sqrt{N}
 \begin{pmatrix}
 \sqrt{1-x}\cos(\theta) \\     
 \sqrt{x}\ex^{i\Delta_2} \\  
 \sqrt{1-x} \sin(\theta)\ex^{i\Delta_3} 
 \end{pmatrix} \,,
 \end{equation}
 where as before $0\leq x \leq 1$. If in contrast the reader is only sensitive to the absolute values, we can effectively describe this situation by neglecting the phases in \eqref{patternVector}, $\Delta_2 = \Delta_3 = 0$

As before, we can discuss states of enhanced memory capacity. Thereby, the role of the negative synaptic connection energy in creating such a state is the same since, up to $1/N$-corrections, the energy expectation value is identical in a number eigenstate and in a coherent state. 
So for $\lambda \ll1$, when the synaptic connection term 
is negligible and the energy of the network is given by the threshold excitation energy, we obtain the energy gap \eqref{DIFFbig}. When we remember that due to the properties of coherent states only patterns 
with large parameter differences that satisfy \eqref{distance} count as distinct patterns, we conclude that the energy difference for 
distinguishable patterns is given by the threshold excitation energy and therefore necessarily large, $\Delta E  \gtrsim 1$.   
 This situation changes dramatically in the critical state, where a stationary inflection point appears in the energy function $H_{\text{bog}}(\vec{a}, \vec{a}^*)$. 
 	Because of the corresponding flat direction,  the energy difference between distinct patterns collapses to zero, as is evident from \eqref{DIFFsmall}.  Hence the system can store 
 different	patterns within an arbitrarily narrow energy gap.

In order to attain such a critical state of enhanced memory storage, one has to proceed as follows when one is given the system \eqref{HNET} with some small coupling $\alpha\ll 1$. First, one needs to go to a total excitation level of $N = \lambda_{lm}/\alpha \approx 1.8/\alpha$. Then one has to distribute those excitations so that the expectation values in the three neurons approximately match the stationary inflection point of the Bogoliubov Hamiltonian. This means one has to choose $\braket{\hat{a}_2^\dagger \hat{a}_2} = x_{\text{inf}} N \approx 0.32 N$ as well as  $\braket{\hat{a}_1^\dagger \hat{a}_1} = (1-x_{\text{inf}})\cos^2(\theta_{\text{inf}}) N \approx 0.67 N$ and $\braket{\hat{a}_3^\dagger \hat{a}_3} = (1-x_{\text{inf}})\sin^2(\theta_{\text{inf}}) N \approx 0.01 N$. Around such a state, an increased number of patterns exists in a small energy gap, provided $N$ is big enough.
	
One can also repeat this procedure for different, i.e., non-critical, values of $\lambda$. Then one finds that the energy gap is not small, $\Delta E \gtrsim 1$. Determining in this way the minimal energy for pattern storage as a function of $\lambda$, one reproduces \fig \ref{fig:bogoGap}. Of course, it is important to note that this plot is only valid in the limit of infinite $N$. For finite $N$, corrections appear that scale as a power of $1/N$. In particular, the critical value of $\lambda$, at which the enhanced memory storage takes place, deviates slightly from $\lambda_{lm}$. In \fig \ref{fig:lambdaVsN}, this critical value of $\lambda$ is shown for some exemplary finite excitation levels $N$.

A small energy gap has a direct implication on the longevity of states, i.e., on the timescale $t_{\text{coh}}$ on which excitation levels of the neurons start to change. As is exemplified in \fig \ref{fig:slowModesTime}, this timescale is short, $t_{\text{coh}} \approx 1$, for states away from the critical point (\fig \ref{fig:slowModesTimea} and \ref{fig:slowModesTimec}). In contrast, it is long, $t_{\text{coh}} \gg 1$, for critical states (\fig \ref{fig:slowModesTimeb}). If one investigates this timescale of evolution for different values of $\lambda$, this leads to \fig \ref{fig:meanFrequency}. We observe that states that evolve significantly more slowly appear at the critical point.
	
The mechanism for memory storage, which is based on assisted gaplessness, can be summarized as follows. In the presence of excitatory synaptic connections, we increase the total excitation level to a point at which a flat direction appears in the energy landscape. On this plateau, there exists a large number of distinct states within a small energy gap. Those states are, however, very close, so typically one would expect that they mix very quickly and information gets washed out. But the key point is that we can distinguish them because they evolve very slowly. Thus, if we read out a state on a timescale smaller than its timescale of evolution, we will encounter it precisely as we have put it in the system.

   To conclude this section, we have seen that 
   a simple system of cold bosons truncated only to three modes effectively describes a neural network 
   of remarkable complexity. Most importantly,
once excited to a critical level, it forms states of sharply enhanced memory capacity in which a large number of patterns can be stored within an 
arbitrarily  very narrow energy gap. This behavior also persists when taking the classical limit of the neural network.
   The above connection opens up the possibility of simulating 
  enhanced memory capacity neural networks in 
  laboratory experiments with cold bosons.

\section{Summary}
 \label{sec:outlook}
 
   In this paper, we have studied 
a phenomenon which we call 
{\it assisted gaplessness}. Reduced to its bare 
essentials, it can be described as a situation in which some highly-excited "master"  degrees of freedom assist others -- connected with the master modes via negative energy couplings -- in becoming {\it effectively gapless} \cite{neural1,neural2}.  In this way, gapless degrees of freedom can emerge within a compact physical system.

The importance of these nearly-gapless modes lies in the fact that they increase the number of energetically-degenerate microstates. Consequently, a large amount of information can be stored within a small energy gap. Moreover, their time evolution becomes slower and therefore the decoherence time increases. This effect becomes more pronounced for a large excitation $N$ of the master modes, i.e., the gap collapses to zero in this limit and the decoherence time diverges. In this way, such critical states become ideal storers of quantum information. 

  Thus, the above phenomenon has a number of important implications. 
First, it can serve as an effective toy model for black hole microstate entropy and holography, 
as it was originally hypothesized in \cite{NPortrait,quantumPhase}. 
This motivation is particularly strengthened by
the model of \cite{areaLaw}, in which the
assisted gaplessness explicitly gives rise to
holographic modes with their number scaling as the area of a lower dimensional sphere and a corresponding microstate entropy  
of  similar scaling, in striking similarity to the Bekenstein entropy of a black hole.  
 
 Moreover, the assisted gaplessness mechanism leads to states of exponentially enhanced pattern storage capacity in neural networks \cite{neural1,neural2}.  On the one hand, this fact could hint towards a possible unity of enhanced memory states 
in neural networks and in gravitational and quantum field theoretic systems. On the other hand, it may allow to understand deep quantum field theoretic (and quantum gravity) concepts, such as black hole entropy and holography, in the language of neural networks. 
In particular, one could use neural networks as a toy laboratory 
 for understanding gravity.\footnote
 {Another important aspect 
 that will not be discussed here is the understanding of quantum transitions to 
classical states of enhanced memory capacity
(such as black hole creation in the collision of elementary particles)  in the language of neural networks \cite{neural3}.} 

In the above light, two important tasks are
to come up with universal methods for identifying states of enhanced memory capacity in generic systems 
of interacting degrees of freedom  and to prepare a 
basis for a possible experimental realizations of such systems. In this paper, we would like to contribute to achieving theses objectives. In particular, we have developed a procedure, which we call $c$-number method, for finding critical states with gapless modes and enhanced storage capabilities in a generic system of attractive bosons.

 We have applied our $c$-number method to a prototype model, which we obtained from truncating a system of attractive cold bosons contained in a one-dimensional box with Dirichlet boundary conditions. Our method enabled us to predict that at a certain critical value of the collective coupling, states emerge in which a gapless collective mode leads to a sharply enhanced memory capacity. We confirmed their existence by numerical analysis. In particular, we were able to observe a distinct increase of the decoherence time already for relatively small values of $N$.

That assisted gaplessness already occurs in such a simple system raises the hope that its experimental realization is feasible. This could e.g., be achieved in experiments with cold atoms. If one can bring them to the critical point, then information could be encoded and retrieved in them by means of soft external radiation. On the one hand, such experiments would be interesting from a general perspective of quantum information. On the other hand, they could provide a framework to simulate other systems of enhanced memory capacity, such as black holes and neural networks.

\section*{Acknowledgements}
 We are grateful to Daniel Flassig for kindly providing us with the Mathematica-files used in \cite{daniel, nico} and to Mischa Panchenko for collaboration in early stages of the project. We thank C\'{e}sar G\'{o}mez for interesting discussions. The work of G.D. was supported by the Humboldt Foundation under Alexander von Humboldt Professorship, the ERC Advanced Grant  "Selfcompletion"  (Grant No. 339169), FPA 2009-07908, CPAN (CSD2007-00042), HEPHACOSP-ESP00346, and by TR 33 "The Dark Universe". 

\appendix
\section{Formulas}
\label{app:formulas}
\begin{itemize}
	\item First and second derivative of Bogoliubov Hamiltonian \eqref{macroscopicHamiltonian} for $\Delta_2=\Delta_3=0$:
\begin{subequations} \label{macroscopicHamiltonianFirstOrder}
	\begin{align*}
	\frac{1}{N} \frac{\partial H_{\text{bog}}}{\partial x} &=  \frac{1}{16} \Big[-16 \lambda  \sin (2 \theta )-2 \lambda  \sin (4 \theta )+16 \cos (2 \theta )-9 \lambda +28 \lambda  x \sin (2 \theta )
	\\
	&+2 \lambda  x \sin (4 \theta )+3 \lambda  (x-1) \cos (4 \theta )+21 \lambda 
	x-4\Big] \,, \numberthis 
	\\
	\frac{1}{N} \frac{\partial H_{\text{bog}}}{\partial \theta} &= \frac{1}{4}(x-1) \Big[-8 \sin (2 \theta )+\lambda  (x-1) \cos (4 \theta )
	\\
	&-\lambda  \cos (2 \theta ) \big(3 (x-1) \sin (2 \theta )
	 -7 x+1\big)\Big]\,, \numberthis 
	 \end{align*}
	 \end{subequations}
 \vspace{-1.3\baselineskip}
	 \begin{subequations}  \label{macroscopicHamiltonianSecondOrder}
\begin{align*}
	\frac{1}{N}\frac{\partial^2 H_{\text{bog}}}{\partial x^2} &= \frac{1}{16} \lambda  \big(28 \sin (2 \theta )+2 \sin (4 \theta )+3 \cos (4 \theta )+21\big) \,,
	\numberthis \\
	\frac{1}{N}\frac{\partial^2 H_{\text{bog}}}{\partial x \, \partial \theta} &= \frac{1}{4} \big(-8 \sin (2 \theta )+2 \lambda  (7 x-4) \cos (2 \theta )\\
	&+\lambda  (x-1) (2 \cos (4 \theta )-3 \sin (4 \theta ))\big) \,,
\numberthis 	\\
	\frac{1}{N}\frac{\partial^2 H_{\text{bog}}}{\partial \theta^2} &= \frac{1}{2} (1-x) \Big(8 \cos (2 \theta )+\lambda  \big((7 x-1) \sin (2 \theta )+2 (x-1) \sin (4 \theta )
	\\
	&+3 (x-1) \cos (4 \theta )\big)\Big) \,.\numberthis 
	\end{align*}
\end{subequations}

	\item Second-order expansion of the  full quantum Hamiltonian \eqref{truncatedHamiltonian} around the point defined by the replacements \eqref{macroscopicReplacements} of macroscopic occupation for $\Delta_2 = \Delta_3 = 0$:
	\begin{equation}
	H_{\text{quad}} =  H_{\text{quad}}^{(1)} + \frac{1}{2} H_{\text{quad}}^{(2)} \,,
	\label{eq1}
	\end{equation}
	where we neglected the constant zeroth order. The first and second order are given by
	\begin{align*}
\frac{H_{\text{quad}}^{(1)}}{\sqrt{N}} &= \frac{1}{8 \sqrt{(1-x) \cos ^2(\theta )}} \bigg[ 
	\\
	& +6 \hat{a}_2 \sqrt{x} \Big(3 \lambda  (1-x)^{3/2} \sin ^3(\theta )+\lambda  \sqrt{1-x} (4 x-3) \sin (\theta )
	\\
	& +(\lambda  (2 x-1)+1) \sqrt{-(x-1) \cos ^2(\theta )}+\lambda  \tan ^2(\theta ) \big((1-x) \cos ^2(\theta
	)\big)^{3/2}\Big)
	\\
	&+\hat{a}_3 \Big(\lambda  (x-1)^2 \cos (4 \theta )+\lambda  (7 x-1) (x-1) \cos (2 \theta )
	\\
	&+\sqrt{4-4 x} \sin (\theta ) \sqrt{-(x-1) \cos ^2(\theta )} (3 \lambda  (x-1) \cos (2 \theta )+8)\Big)\bigg]
	\\
	&+ \text{h.c.} \numberthis  \label{quantumHamiltonianFirstOrder}
\intertext{and}
	H_{\text{quad}}^{(2)}&= \frac{1}{128 \left((1-x) \cos ^2(\theta )\right)^{3/2}}\bigg[
	\\
	&+16 \hat{a}_2 \hat{a}_2 \lambda\bigg(  \sqrt{1-x} ((23-16 x) x-4) \sin (\theta )
	\\
	&+4 (4 x-1)  \left((1-x) \cos ^2(\theta )\right)^{3/2}
	\\
	&-(1-x)^{3/2} \sin
	^3(\theta ) (2 (x-1) \cos (2 \theta )+21 x-6)\bigg)
	\\
	&+ 16 \hat{a}_2^\dagger \hat{a}_2 \bigg(2 \sec ^2(\theta ) (\lambda  (10 x-1)+3) \big((1-x) \cos ^2(\theta )\big)^{3/2}
	\\
	& +\sin (\theta ) \Big(-14 \lambda  (1-x)^{5/2} \sin ^4(\theta )-7 \lambda  (1-x)^{3/2} (7 x-4) \sin ^2(\theta )
	\\
	& -6 \lambda  \tan
	^3(\theta ) \sec (\theta ) \left(-(x-1) \cos ^2(\theta )\right)^{5/2} +\lambda  \sqrt{1-x} ((49-32 x) x-14)
	\\
	&-2 \tan (\theta ) \sec (\theta ) (\lambda  (13 x-4)+3) \big((1-x) \cos ^2(\theta )\big)^{3/2}\Big)\bigg)
	\\
	&+16\hat{a}_2 \hat{a}_3  \lambda  \sqrt{x} \big((1-x) \cos ^2(\theta )\big) \bigg(8 (x-1) \cos (2 \theta )+3 x \sec ^2(\theta )\\
	& +10 \sqrt{1-x} \sin (\theta ) \sqrt{-(x-1) \cos ^2(\theta )}-x+1\bigg)
	\\
	&+16\hat{a}_2 \hat{a}_3^\dagger \lambda  \sqrt{x}\big((1-x) \cos ^2(\theta )\big) \bigg(10 (x-1) \cos (2 \theta )+3 x \sec ^2(\theta )
	\\
	& +2 \sqrt{1-x} \sin (\theta ) \sqrt{-(x-1) \cos ^2(\theta )}+x-1\bigg)
	\\
	&+\hat{a}_3\hat{a}_3 \lambda  (x-1)\bigg( 32 \cos (2 \theta ) \sqrt{(1-x) \cos ^2(\theta )}+32 x \sqrt{(1-x) \cos ^2(\theta )}
	\\
	&+32 \cos (4 \theta ) \sec ^2(\theta ) \big((1-x) \cos ^2(\theta )\big)^{3/2}
	\\
	&\ -6 \sqrt{1-x} \sin (\theta )
	(4 (3 x-2) \cos (2 \theta )+3 (x-1) \cos (4 \theta )+17 x-5)\bigg)
	\\
	&+16\hat{a}_3 \hat{a}_3^\dagger \big((1-x) \cos ^2(\theta )\big)\sec ^2(\theta ) \bigg(2 (\lambda  (3 x-1)+8) \sqrt{(1-x) \cos ^2(\theta )}
	\\
	&+\sin (\theta ) \Big(\sin (\theta ) \Big(\lambda  \sin (\theta ) \Big(5 (x-1) \sin (\theta ) \Big(3 \sqrt{1-x} \sin (\theta )
	\\
	&+4 \sqrt{(1-x) \cos ^2(\theta )}\Big)
	+9 \sqrt{1-x} (3-4 x)\Big)
	\\
	&-2 (\lambda  (13 x-11)+8) \sqrt{(1-x) \cos ^2(\theta )}\Big)+12 \lambda  \sqrt{1-x} (2 x-1)\Big)\bigg)
	\\
	&+\text{h.c.} \,.\numberthis  \label{quantumHamiltonianSecondOrder}
	\end{align*}
	
\end{itemize}

\section{Example: Review of Periodic Bose Gas}
\label{app:periodic}
\subsection{Application of $C$-Number Method}
In order to provide a detailed example, we apply the $c$-number method to the one-dimensional Bose gas with periodic boundary conditions. For the repulsive case, in which we are not interested,  it goes under the name of  Lieb-Liniger model \cite{LL}.  The attractive case was considered previously in \cite{kanamoto}. Its quantum information features have already been studied in a series of papers \cite{quantumPhase, daniel, nico, goldstone,mischa1,mischa2,mischa3}. In particular, the replacement of the Hamiltonian by a $c$-number function has already been used to find its critical point in \cite{goldstone}. Except for the boundary conditions, this system is the same as our prototype model studied in  section \ref{sec:3Mode}.
Its Hamiltonian is therefore identical to \eqref{periodicHamiltonian}, but the free eigenfunctions change according to the periodic boundary conditions:
\begin{equation}
\hat{\psi} = \sqrt{\frac{1}{L}}\sum_{k=-\infty}^\infty \hat{a}_k \exp\left(\frac{2\pi i k  z}{L}\right) \,.
\label{periodicExpansion}
\end{equation}
Consequently, we obtain the Hamiltonian in momentum space:
\begin{align}
\centering
\hat{H} = & \frac{4\pi^2\hbar^2}{2m L^2} \bigg[\sum_{k=-\infty}^\infty  k^2 \hat{a}_k^\dagger \hat{a}_k
-  \frac{\alpha }{4} \sum_{k,l,m =-\infty}^\infty
\hat{a}_k^\dagger \hat{a}_l^\dagger \hat{a}_{m+k} \hat{a}_{l-m}\bigg] \,.
\label{momentumHamiltonianPeriodic}
\end{align}
For convenience, we set $L = 2\pi$ and $\hbar = 2m = 1$ from now on.

For small coupling, we expect a regime where momentum modes with $|k|>1$ are suppressed due to their higher kinetic energy. Therefore, we first truncate \eqref{momentumHamiltonianPeriodic} to the modes with $|k|\leq 1$. This assumption will be justified later.  Following the method introduced in section \ref{sec:analysis}, we then replace the creation and annihilation operators by $c$-numbers. As explained, we have to take into account symmetries in this procedure.
The first one is conservation of particle number, which is incorporated in the replacement rule of the $0$-mode. An additional symmetry of the system consists in momentum conservation. In the superselection sector of zero momentum, it implies that the expectation values of the particle numbers in the $1$- and $-1$-mode have to be the same: $\braket{a_1^\dagger a_1} = \braket{a_{-1}^\dagger a_{-1}}$. Moreover, the Hamiltonian is invariant under an additional phase symmetry, $\hat{a}_1 \rightarrow \ex^{i \phi}\hat{a}_1 $ and $\hat{a}_{-1} \rightarrow \ex^{- i \phi} \hat{a}_{-1} $. So in total, we have to eliminate an absolute value and a phase. In addition to the replacement rule \eqref{replace0} due to particle number conservation, we consequently get: $a_{-1} \rightarrow a_1$. Thus, the Bogoliubov replacement \eqref{bogoliubovReplacement} reads

\begin{equation}
\hat{\vec{a}} = \begin{pmatrix}
\hat{a}_{-1} \\
\hat{a}_{1}
\end{pmatrix}
\rightarrow \begin{pmatrix}
a_{1} \\
a_{1}
\end{pmatrix}  \, , \qquad \hat{\vec{a}}^\dagger = \begin{pmatrix}
\hat{a}^\dagger_{-1} \\
\hat{a}^\dagger_{1} \end{pmatrix} \rightarrow
\begin{pmatrix}
a_{1}^* \\
a_{1}^*
\end{pmatrix} \,,
\end{equation}
and
\begin{equation}
\hat{a}_0 \rightarrow \sqrt{N - 2 |a_{1}|^2} \, , \qquad \hat{a}_0^\dagger \rightarrow \sqrt{N-2 |a_{1}|^2} \, .
\end{equation}
This gives the following Bogoliubov Hamiltonian:
 \begin{equation}
H_{\text{bog}} = 2|a_{1}|^2  -\frac{\alpha}{4} 
\bigg (N^2 + 2 N ( a_1  +  a_1^*)^2-2|a_1|^2(3|a_1|^2 + 2 a_1^2  +2  a_1^{*2})\bigg) \,.
\end{equation}

For this so obtained complex-valued function, we want to find a flat direction. First we look for an extremal point by finding a solution to \eqref{extremum}:
	\begin{align}
	\label{eq:periodicFirstDer}
	\frac{\partial H_{\text{bog}}}{\partial a_1}   =  
	2 a_1^* - \alpha \left(N (a_1 + a_1^*) -3 a_1^2 a_1^* - 3 a_1 a_1^{*2} - a_1^{*3}\right) = 0 \,.
	\end{align}
 An obvious solution is $a_1 = 0$.\footnote
 {There are two other solutions at $a_1 \approx \pm 0.5345 \sqrt{\frac{\alpha N -1}{\alpha}}$. However, the determinant of the second derivative matrix $\mathcal{M}$  never vanishes at these points, i.e., there is no flat direction.} 
 The second step is to evaluate the matrix $\mathcal{M}$ of second derivatives at this point. We need to determine when it fulfills \eqref{inflection}, i.e., when its determinant vanishes:
	\begin{equation}
\det \mathcal{M} = -4 + 4 \alpha N = 0 \,.
	\end{equation} 
This is the case for a collective coupling $\lambda =\alpha N =1$. Therefore, we expect a critical mode to appear for $\lambda_{lm} = 1$ in a state where all particles are in the $0$-mode.

\subsection{Accordance with Previous Findings}
\label{ssec:previous}
In the following, we want to confirm the existence of this critical point by reviewing the findings previously obtained in  \cite{kanamoto, quantumPhase, daniel, nico, goldstone,mischa1,mischa2,mischa3}. First, still working in the limit of large $N$, we directly determine the spectrum of quantum fluctuations. Thus, we expand the Hamiltonian \eqref{momentumHamiltonianPeriodic} to second order in creation and annihilation operators around the point defined by the Bogoliubov approximation. Focusing on states in which only the 0-mode is occupied, $\hat{a}_0 \rightarrow \sqrt{N}$, we therefore obtain:
\begin{equation}
\hat{H} = \sum_{k \ne 0} \Big(k^2- \frac{\lambda}{2}\Big)\hat{a}_k^\dagger \hat{a}_k - \frac{1}{4} \lambda \sum_{k \ne 0} (\hat{a}_k^\dagger \hat{a}_{-k}^\dagger + \hat{a}_k \hat{a}_{-k}) \,,
\end{equation}
with the collective coupling $\lambda = \alpha N$. Now we perform the Bogoliubov transformation
\begin{equation}
\hat{a}_k = u_k \hat{b}_k + v_k^*\hat{b}_{-k}^\dagger \,,
\end{equation}
where we have 
\begin{equation}
u_k^2 = \frac{1}{2}\left ( 1+\frac{k^2-\frac{\lambda}{2}}{\epsilon_k}\right) \,, \qquad
v_k^2 = \frac{1}{2} \left (\frac{k^2-\frac{\lambda}{2}}{\epsilon_k} -1\right)\,.
\end{equation}
We obtain the diagonalized Hamiltonian
\begin{equation}
\hat{H} = \sum_{k \ne 0} \epsilon_k \hat{b}_k^\dagger \hat{b}_k\,, \qquad \epsilon_k = \sqrt{k^2(k^2-\lambda)} \,. \label{bogoliubovHamiltonian}
\end{equation}
This analysis confirms that a gapless mode exists at $\lambda_{\text{lm}} = 1$.

\begin{figure}
	\centering 
	\begin{subfigure}{0.45\textwidth}
		\includegraphics[width=\textwidth]{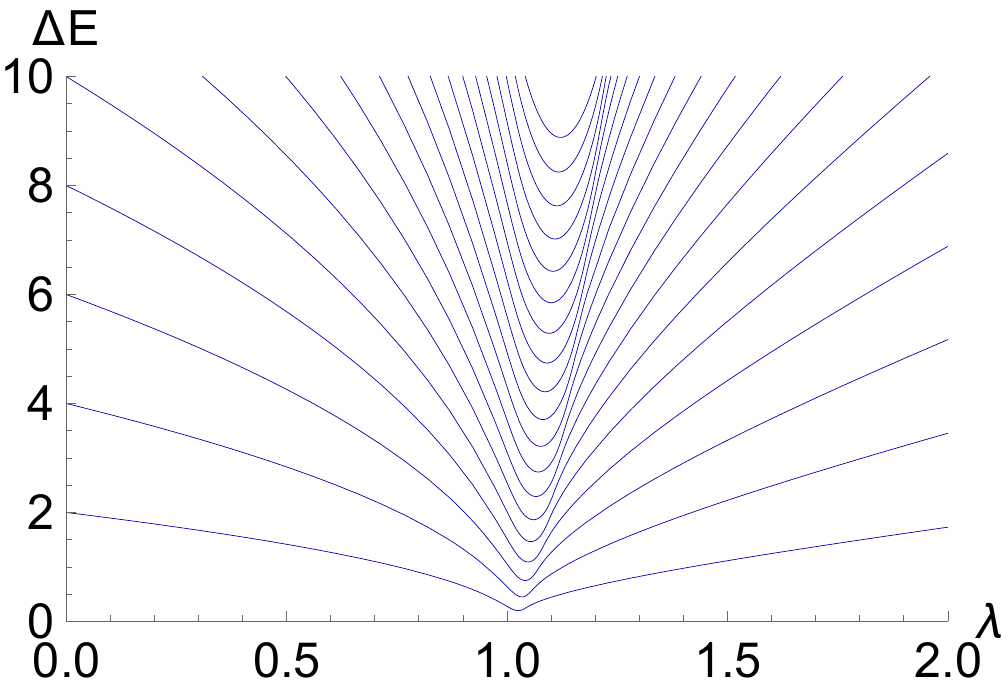}
		\caption{ Energies of the lowest-lying excitations as functions of $\lambda$. They become nearly-gapless at $\lambda = \lambda_{\text{lm}}=1$.}
		\label{fig:periodicExcitationGaps}
	\end{subfigure}
		\hspace{0.05\textwidth}
	\begin{subfigure}{0.45\textwidth}
	\includegraphics[width=\textwidth]{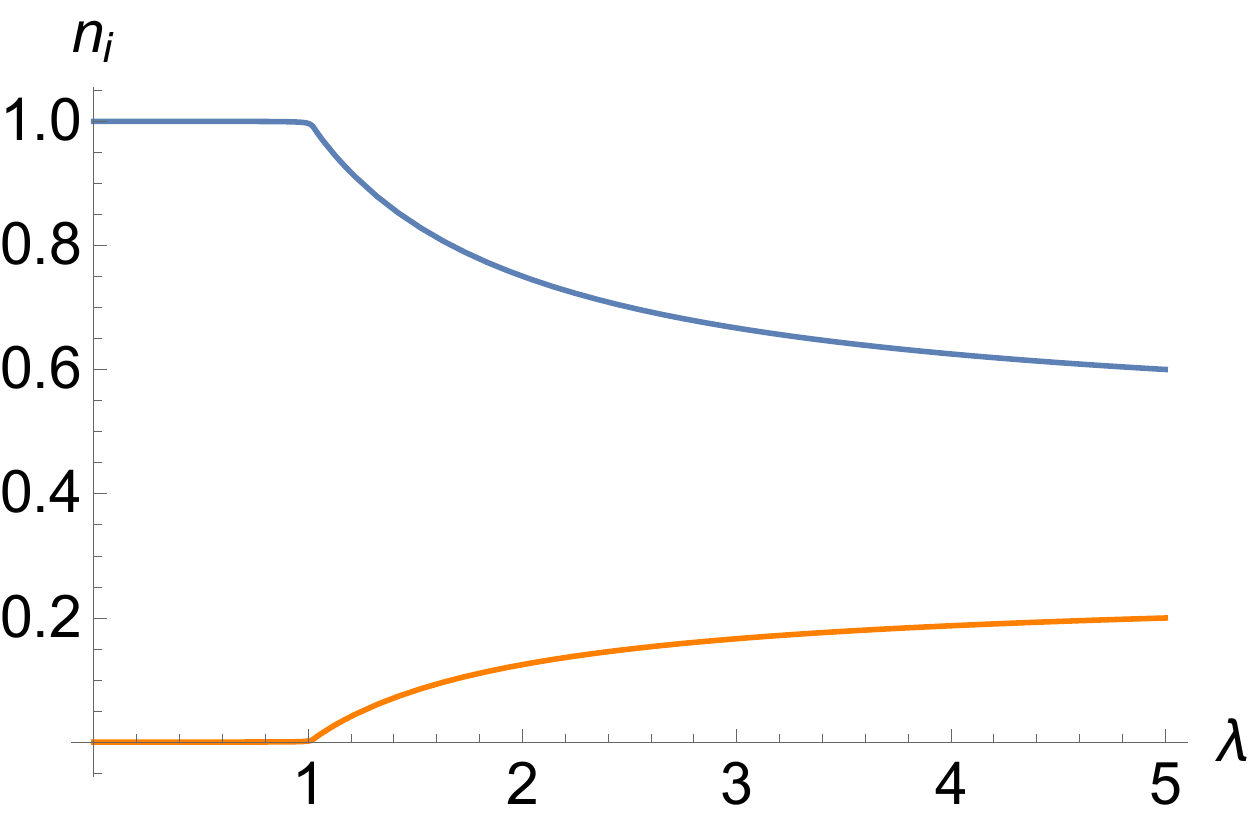}
	\caption{Relative occupation numbers of the ground state as functions of $\lambda$. The $0$-mode is displayed in blue and the $\pm1$-modes in orange. The occupation numbers start to change continuously for $\lambda > \lambda_{\text{lm}}=1$.}
	\label{fig:periodicGroundStateOccupation}
	\end{subfigure}
	\caption{Numerical analysis of the periodic system for $N=1000$. The Hamiltonian \eqref{momentumHamiltonianPeriodic} is truncated to the $0$- and $\pm1$-modes. At $\lambda_{\text{lm}}=1$, a gapless mode appears in the spectrum and the ground state undergoes a second-order phase transition.}
\end{figure}

Finally, we also study the system at finite $N$. We do so by numerically diagonalizing the full Hamiltonian.
Then we obtain the spectrum of fluctuations as the energy gap between the lowest-lying eigenstates. \fig \ref{fig:periodicExcitationGaps} shows that also this analysis leads to a critical point at $\lambda_{lm}=1$. Of course, corrections appear for finite $N$ which scale as a power of $1/N$: The gap scales as $1/N^{1/3}$ and moreover, the critical value of $\lambda$ is no longer at $1$, where the deviation scales as $1/N^{2/3}$ \cite{kanamoto}. 

For the system with periodic boundary conditions, the critical point $\lambda_{lm}=1$ has a second meaning. As is evident from \fig \ref{fig:periodicGroundStateOccupation}, it corresponds to a second-order phase transition of the ground state: For $\lambda < \lambda_{lm}$, only the $0$-mode is macroscopically occupied in the ground states whereas also the other modes get populated for $\lambda > \lambda_{lm}$. Since the occupation numbers of the ground state change continuously, higher modes get populated slowly for $\lambda \gtrsim \lambda_{\text{lm}}$. In this regime, it is therefore possible to truncate the system to the $0$- and $\pm1$-modes, as was shown explicitly in \cite{daniel}. This allows for a straightforward numerical analysis.

One can also analyze the generation of entanglement in this system. For an initial state in which only the $0$-mode is occupied, it was shown the time required to generate one-particle entanglement is long ($\sim N$) for $\lambda < \lambda_{\text{lm}}$ and short ($\sim \ln N$) for $\lambda > \lambda_{\text{lm}}$ \cite{nico, mischa1}. By coupling the quantum gas to an external system, one can moreover use it to perform quantum computing \cite{mischa1, mischa3}.

\providecommand{\href}[2]{#2}\begingroup\raggedright\endgroup


\begin{thebibliography}{10}
	
	\bibitem{neural1}
G.~Dvali, {\em Critically excited states with enhanced memory and pattern recognition capacities in quantum brain networks: Lesson from black holes\/},  \href{http://arxiv.org/abs/1711.09079}{{\tt arXiv:1711.09079
		[quant-ph]}}.
	
			\bibitem{neural2}
				G.~Dvali, {\em Black Holes as Brains: Neural Networks with Area Law Entropy\/}, 
				\href{https://doi.org/10.1002/prop.201800007}{Fortschr. Phys. {\bf 66} (2018) no.~4, 1800007},
			\href{http://arxiv.org/abs/1801.03918}{{\tt arXiv:1801.03918
						[hep-th]}}.
	
	\bibitem{quantumPhase}
	G.~Dvali and C.~Gomez, {\em {Black Holes as Critical Point of Quantum Phase
			Transition}\/},  \href{http://dx.doi.org/10.1140/epjc/s10052-014-2752-3}{Eur.
		Phys. J. C {\bf 74} (2014)  2752},
	\href{http://arxiv.org/abs/1207.4059}{{\tt arXiv:1207.4059 [hep-th]}}.
	
			\bibitem{areaLaw}
				G.~Dvali, {\em Area Law Micro-State Entropy from Criticality and Spherical
				Symmetry\/},  \href{https://doi.org/10.1103/PhysRevD.97.105005} {Phys. Rev. D. {\bf 97} (2018) no.~10, 105005},
			\href{http://arxiv.org/abs/1712.02233}{{\tt arXiv:1712.02233
						[hep-th]}}.

	
		\bibitem{goldstone}
	G.~Dvali, A.~Franca, C.~Gomez, and N.~Wintergerst, {\em Nambu-Goldstone
		Effective Theory of Information at Quantum Criticality\/},
	\href{http://dx.doi.org/10.1103/PhysRevD.92.125002}{Phys. Rev. D {\bf 92}
		(2015) no.~12, 125002}, \href{http://arxiv.org/abs/1507.02948}{{\tt
			arXiv:1507.02948 [hep-th]}}.
	

				
		
		
		\bibitem{mischa1}
		G.~Dvali and M.~Panchenko, {\em Black Hole Type Quantum Computing in Critical
			Bose-Einstein Systems\/},  \href{http://arxiv.org/abs/1507.08952}{{\tt
				arXiv:1507.08952 [hep-th]}}.
		
		
		\bibitem{mischa3}
		G.~Dvali and M.~Panchenko, {\em Black Hole Based Quantum Computing in Labs and
			in the Sky\/},  \href{http://dx.doi.org/10.1002/prop.201600060}{Fortschr.
			Phys. {\bf 64} (2016) no.~8-9, 569--580},
		\href{http://arxiv.org/abs/1601.01329}{{\tt arXiv:1601.01329 [hep-th]}}.
	
	\bibitem{daniel}
	D.~Flassig, A.~Pritzel, and N.~Wintergerst, {\em {Black Holes and Quantumness
			on Macroscopic Scales}\/},
	\href{http://dx.doi.org/10.1103/PhysRevD.87.084007}{Phys. Rev. D {\bf 87}
		(2013) no.~8, 084007},
	\href{http://arxiv.org/abs/1212.3344}{{\tt arXiv:1212.3344 [hep-th]}}.


		\bibitem{nico}
	G.~Dvali, D.~Flassig, C.~Gomez, A.~Pritzel, and N.~Wintergerst, {\em
		{Scrambling in the Black Hole Portrait}\/},
	\href{http://dx.doi.org/10.1103/PhysRevD.88.124041}{Phys. Rev. D {\bf 88}
		(2013) no.~12, 124041},
	\href{http://arxiv.org/abs/1307.3458}{{\tt arXiv:1307.3458 [hep-th]}}.

	
	\bibitem{bogoliubov}
	N.~Bogoliubov, {\em On the theory of superfluidity\/},  J. Phys. {\bf 11}
	(1947) no.~1, 23--32.
	
		
	
		
			\bibitem{bekensteinEntropy}
		J.~D. Bekenstein, {\em Black holes and entropy\/},
		\href{http://dx.doi.org/10.1103/PhysRevD.7.2333}{Phys. Rev. D {\bf 7} (1973)
			no.~8, 2333--2346}.	
		
			
	\bibitem{bekensteinBound}
			J.~D. Bekenstein, {\em Universal upper bound on the entropy-to-energy ratio for
				bounded systems\/},  \href{http://dx.doi.org/10.1103/PhysRevD.23.287}{Phys.
				Rev. D {\bf 23} (1981) no.~2, 287--298}. 
				

	


	
	\bibitem{bremermannBoud}
	H.~J. Bremermann, {\em Quantum noise and information\/},  in {\em Proceedings
		of the Fifth Berkeley Symposium on Mathematical Statistics and Probability} {\bf 4} (1967), 15--20.


	
\bibitem{NPortrait}
G.~Dvali and C.~Gomez, {\em {Black Hole's Quantum N-Portrait}\/},
\href{http://dx.doi.org/10.1002/prop.201300001}{Fortschr. Phys. {\bf 61}
	(2013)  742--767},
\href{http://arxiv.org/abs/1112.3359}{{\tt arXiv:1112.3359 [hep-th]}}.

	
	\bibitem{kanamoto}
	R.~Kanamoto, H.~Saito, and M.~Ueda, {\em Quantum phase transition in
		one-dimensional Bose-Einstein condensates with attractive interactions\/},
	\href{https://dx.doi.org/10.1103/PhysRevA.67.013608}{Phys. Rev. A {\bf 67}
		(2003) no.~1, 013608}, \href{http://arxiv.org/abs/cond-mat/0210229}{{\tt
			arXiv:cond-mat/0210229}}.



 \bibitem{hologBH} 
 
 G.'t~Hooft, {\em Dimensional reduction in quantum gravity\/}, {\tt Conf.Proc. C930308 (1993), 284-296}, \href{https://arxiv.org/abs/gr-qc/9310026}{{\tt arXiv:gr-qc/9310026}}; \\
 L.~Susskind,
{\em The World As A Hologram\/}, \href{https://doi.org/10.1063/1.531249}{J. Math. Phys. {\bf 36}  (1995), 6377}, 
\href{https://arxiv.org/abs/hep-th/9409089}{{\tt arXiv:hep-th/9409089}}.

			
\bibitem{hologAdS}
E.~Witten, {\em Anti-de Sitter space and holography\/},
\href{https://doi.org/10.4310/ATMP.1998.v2.n2.a2}{ Adv. Theor. Math. Phys. {\bf 2} (1998), 253 }, \href{https://arxiv.org/abs/hep-th/9802150}{{\tt arXiv:hep-th/9802150}};\\
L.~Susskind and E. Witten, {\em The holographic bound in anti-de Sitter space\/},
\href{https://arxiv.org/abs/hep-th/9805114}{{\tt arXiv:hep-th/9805114}}.





	
 \bibitem{experimental}
 I.~Bloch, J.~Dalibard, and S.~Nascimb\`{e}ne, {\em Quantum simulations with ultracold quantum gases\/},
 \href{https://doi.org/10.1038/nphys2259}{ Nature Phys. {\bf 8} (2012), 267};\\
 I.~Bloch, J.~Dalibard, W.~Zwerger, {\em Many-Body Physics with Ultracold Gases\/},
 \href{https://doi.org/10.1103/RevModPhys.80.885}{ Rev. Mod. Phys. {\bf 80} (2008), 885}, 
 \href{https://arxiv.org/abs/0704.3011}{{\tt arXiv:0704.3011 [cond-mat.other]}}.
	
	\bibitem{bogoliubovCalculation1}
	C.~Tsallis, {\em Diagonalization methods for the general bilinear Hamiltonian
		of an assembly of bosons\/},  \href{http://dx.doi.org/10.1063/1.523549}{J.
		Math. Phys. {\bf 19} (1978) no.~1, 277--286}.
		
\bibitem{grossPitaevskii}
E.P.~Gross, {\em Structure of a quantized vortex in boson systems\/},
\href{https://doi.org/10.1007\%2FBF02731494}{Il Nuovo Cimento {\bf 20} (1961) no.~3, 454--457};\\
L.P.~Pitaevskii, {\em Vortex lines in an imperfect Bose gas},
Sov. Phys. JETP. {\bf 13} (1961) no.~2., 451--454.		
		
	\bibitem{mathematica}
	{Wolfram Research, Inc.}, {\em Mathematica 11.1.1\/},  2017.
	
\bibitem{bogoliubovCalculation2}
Y.~Tikochinsky, {\em On the diagonalization of the general quadratic
	Hamiltonian for coupled harmonic oscillators\/},
\href{http://dx.doi.org/10.1063/1.524093}{J. Math. Phys. {\bf 20} (1979)
	no.~3, 406--408}.

	\bibitem{neural3}
		G.~Dvali, {\em Classicalization Clearly: Quantum Transition into States of Maximal Memory Storage Capacity\/},  \href{https://arxiv.org/abs/1804.06154}{{\tt 	arXiv:1804.06154 [hep-th]}}.
				
			\bibitem{LL}
		E.~H. Lieb and W.~Liniger, {\em Exact Analysis of an Interacting Bose Gas. I. The General Solution and the Ground State\/},
		\href{http://dx.doi.org/10.1103/PhysRev.130.1605}{Phys. Rev. {\bf 130} (1963) no.~4, 1605--1616};\\
		E.~H. Lieb, {\em Exact Analysis of an Interacting Bose Gas. {II}. The
			Excitation Spectrum\/},
		\href{http://dx.doi.org/10.1103/PhysRev.130.1616}{Phys. Rev. {\bf 130} (1963)
			no.~4, 1616--1624}.
		
			\bibitem{mischa2}
		M.~Panchenko, {\em The Lieb-Liniger model at the critical point as toy model
			for Black Holes\/},  \href{http://arxiv.org/abs/1510.04535}{{\tt
				arXiv:1510.04535 [hep-th]}}.
			

\end{thebibliography}
\end{document}